\documentstyle[seceq,preprint,epsf]{jpsj}
\newcommand{\ds}{\displaystyle}
\newcommand{\noi}{\noindent}
\newcommand{\non}{\nonumber}

\newcommand{\ii}{{\rm i}}
\newcommand{\kpara}{k_{\parallel}}
\newcommand{\kperp}{k_{\perp}}

\newcommand{\para}{\parallel}
\newcommand{\beq}{\begin{eqnarray}}
\newcommand{\eeq}{\end{eqnarray}}

\newcommand{\be}{\begin{eqnarray}}
\newcommand{\en}{\end{eqnarray}}
\newcommand{\JPSJ}{J. Phys. Soc. Jpn.}

\newcommand{\PRL}{ Phys. Rev. Lett.}
\newcommand{\PRB}{ Phys. Rev.{\bf  B}}

\newcommand{\PROG}{Prog. Theoret. Phys}

\def\runtitle{Dimensional Crossovers in the Doped Ladder System}
\def\runauthor{Jun-ichiro Kishine and Kenji Yonemitsu}
\title{Dimensional Crossovers in the Doped Ladder System:
Spin Gap, Superconductivity and Interladder Coherent Band Motion} 
\author{Jun-ichiro Kishine\thanks{E-mail:kishine@ims.ac.jp} and Kenji 
Yonemitsu}
\inst
{Department of Theoretical Studies,
Institute for Molecular Science,
Okazaki 444-8585, Japan}
\recdate{December 5, 1997}
\abst
{Based on the perturbative renormalization group (PRG) approach,
we have studied dimensional crossovers in Hubbard ladders coupled via
weak interladder one-particle hopping, $t_{\perp}$.
We found that 
the one-particle crossover is strongly suppressed through
 growth of the intraladder scattering processes which lead the isolated
 Hubbard ladder system toward the spin gap metal (SGM) phase.
Consequently when $t_{\perp}$ sets in, there exists, 
for any finite intraladder Hubbard repulsion, $U>0$, the region where the 
two-particle crossover dominates the one-particle crossover and consequently
the  $d$-wave superconducting transition, which is regarded as a bipolaron condensation,
occurs. 
By solving the scaling equations for the interladder one-particle and
two-particle hopping amplitudes, we   give phase diagrams of the system 
with respect to $U$, $t_{\perp0}$ (initial value
of $t_{\perp}$) and the temperature, $T$.
We compared   the above dimensional crossovers with those in a weakly 
coupled chain system,  clarifying the difference
between them.
}
\kword
{doped  ladder, dimensional crossover, perturbative renormalization-group,
spin gap metal, bipolaron condensation, $d$-wave superconductivity}

\begin{document}
\maketitle
\baselineskip 14pt
\section{Introduction}
Magnetic and electronic properties of  ladder materials have attracted  
great   interest.\cite{DR}
Central to these issues are the   effects of the unusual spin-liquid state 
with a spin excitation gap in the undoped parent system on the electronic 
conduction in
the doped  system.
Last year Uehara {\it et al.}~\cite{AG} discovered a superconductivity 
signal in the doped ladder system, Sr$_{14-x}$Ca$_{x}$Cu$_{24}$O$_{41}$, 
under pressure.
The compound consists of the alternating stacks of   planes with the 
CuO$_{2}$ chains and  with the  Cu$_{2}$O$_{3}$ two-leg ladders.
The nominal valence of Cu is +2.25, independent of $x$, so the system is 
inherently doped with 6 holes in the unit, which
are mainly on the chains at $x=0$.
Increase of the low-energy spectral weight\cite{OMEU} with increasing $x$ 
indicates that the hole carriers are progressively redistributed from 
chains 
to ladders with Ca substitution for Sr.
This feature was also confirmed theoretically through the ionic and 
cluster model calculations.\cite{MTM}
The behavior of the dc resistivity of the compounds  changes from 
semiconducting-like to metal-like with Ca substitution,\cite{AG,MOKEU} 
and  superconductivity sets in for Sr$_{0.4}$Ca$_{13.6}$Cu$_{24}$O$_{41}$ 
below $T_{c}=12$ K under a pressure of 3 GPa.\cite{AG}

The spin excitation gap at the  ladder cite in Sr$_{14}$Cu$_{24}$O$_{41}$  
has been observed through 
NMR shifts and rates,\cite{TKKK,Takigawa}  and neutron scattering 
experiments.\cite{EAT}
A remarkable feature  is that the spin gap in the ladder  survives  upon 
Ca substitution,\cite{KTKK,KM} indicating
the doped ladder compounds, Sr$_{14-x}$Ca$_{x}$Cu$_{24}$O$_{41}$, are in 
the {\it spin gap metal phase} which
reflects the one-dimensional character of the ladder.
According to    the electronic structure calculation of 
Sr$_{14-x}$Ca$_{x}$Cu$_{24}$O$_{41}$ under ambient pressure within the 
local-density approximation,\cite{AT} 
the interladder hoppings  are 5-20 \% of the intraladder ones, indicating 
a pseudo-one-dimensional character of the system.

The superconducting transition  under  pressure suggests that interladder 
one-particle hopping enhanced by applied pressure
plays an important role. 
Recent experiments of the resistivity along the ladders, $\rho_{c}$, of 
the single crystal Sr$_{2.5}$Ca$_{11.5}$Cu$_{24}$O$_{41}$ by Akimitsu {\it 
et al.}~
\cite{MOTT} 
shows that  the  superconductivity  sets in below 10 K under 3.5 GPa 
$\sim$ 8 GPa with the temperature dependence of $\rho_{c}$ changing 
gradually from
$T$-linear to $T^2$.
The ratio of the normal state resistivity, $\rho_{c}/\rho_{a}$ ($\rho_{a}$ 
is resistivity perpendicular to the ladder direction in the ladder plane) 
also indicates a dimensional crossover from 1D to 2D with increasing   
applied pressure.\cite{MOTT}

The low-energy asymptotics of the isolated ladder system  has been 
extensively studied 
mainly based on the  renormalization-group   approach,
with and without the aid of the bosonization technique,   to the 
two-leg Hubbard ladder\cite{MF,FL,KR,BF,Schulz1,NLK,YS1} and the two-leg 
$t$-$J$ 
ladder.\cite{DVK,NN,ST,NO}
In the  isolated Hubbard ladder, the most relevant phase  is characterized 
by a strong coupling fixed point and is denoted by  
{\lq\lq}phase I{\rq\rq} by 
Fabrizio\cite{MF} and {\lq\lq}C1S0 phase{\rq\rq} by Balents and 
Fisher.\cite{BF}
In this phase,   only the total charge mode remains gapless and 
consequently 
the $d$-wave superconducting correlation  becomes the most dominant 
one\cite{MF,KR,BF,Schulz1} and 
the $4k_{F}$-CDW correlation becomes sub-dominant 
one\cite{KR,BF,Schulz1} at least when the intraladder correlation is weak. 
From now on, we call this strong coupling phase a {\lq\lq}spin gap metal 
(SGM) phase{\rq\rq}.
Within the bosonization scheme,  exponents of the $d$-wave 
superconducting   and the $4k_{F}$CDW correlations, $K_{S}$ and $K_{C}$,
satisfy the {\it duality relation}, $K_{S}\cdot K_{C}=1$, suggesting the 
low-energy dynamics in the SGM phase is described with  {\lq\lq}{\it bipolarons}{\rq\rq}, each of which lies
along the rung of the ladder.\cite{NN,TTR1}
The enhancement of the $d$-wave superconducting channel has also been 
found by numerical studies on the two-leg Hubbard ladder
through  the density matrix renormalization group (DMRG) 
analysis,\cite{NWS}
the exact diagonalization study,\cite{HP} and the Monte-Carlo 
simulation.\cite{KKA}
 The exact diagonalization study on the  two-leg $t$-$J$ 
ladder  also gave  evidences of a gapless total charge mode\cite{PSH} and a $d$-wave RVB state\cite{TTR1,KS} in the  
doped system.

So far effects of the weak interladder hopping  have 
been studied through   mean field approximations\cite{TTR2,OG}
and   power counting arguments.\cite{LBF}
Recently the present authors discussed effects 
of the interladder one-particle hopping, $t_{\perp}$, on the low-energy 
asymptotics of the weakly coupled Hubbard ladder system, 
based on the perturbative renormalization-group (PRG) approach and 
discussed 
dimensional crossovers in the system.\cite{KY}

Although the dimensional crossover problem in the coupled ladder system is 
rather new aspect, 
the similar problem  in the coupled chain system has a  history of  about 
twenty years.\cite{Solyom,FPS,Voit}
Effects of strong intrachain quantum fluctuations 
 on the competition between one-particle and
two-particle crossovers were first addressed by Brazovskii and 
Yakovenko.\cite{BY}   
 It was pointed out by Suzumura,\cite{YS} based on the bosonization 
technique,
that the two-particle crossover dominates the one-particle crossover only 
in the presence of large intrachain correlation.
Recently  the competition was again  discussed in terms of the stiffness 
of the Tomonaga-Luttinger liquid.\cite{VMY}

Bourbonnais and Caron, on the other hand, first discussed the problem 
based on the PRG approach,\cite{CB,BC1,BC2} where the intrachain 
interaction
 and interchain one-particle hopping
are treated as perturbations to  one-dimensional free fermion gas.
In their formulation, the two-particle processes are generated by the  
interchain one-particle hopping {\it in the course of renormalization.}
Based on their formulation, the dimensional crossovers in the weakly 
coupled chains were discussed in terms of   
the anomalous dimension of the Tomonaga-Luttinger liquid.\cite{Boies}

In the present paper we    extend discussions on  
the results we shortly presented in  our previous paper.\cite{KY} 
Based on the PRG approach, we treated the {\it intraladder} interaction 
and the {\it interladder} one-particle hopping as perturbations to the 
free fermion gas on the ladder
and obtained the phase diagram of the system.
In this extended version, a full account of application of the PRG scheme 
to the dimensional crossover problem
in the weakly coupled ladder system
is given.
Furthermore,   we re-examine   the dimensional crossovers in the weakly 
coupled chain system,
clarifying
the difference between  the   coupled ladders and chains.
We show that the main difference comes from the  different behaviors of 
the scaling flows of respective
coupling strengths, i.e, the different universality classes 
of the corresponding isolated systems.

The outline of the present paper is as follows.
In  \S 2, we give a qualitative description of the notion of the 
one-particle and two-particle
crossovers in the weakly coupled ladder system.
In  \S 3, we give a full account of formulation; the action,  the scaling 
equations for the 
intraladder processes, the interladder one-particle process and
the interladder two-particle process.
In  \S 4, based on the solutions of the scaling equations, 
we give phase diagrams of the system. 
In  \S 5, we re-examine the dimensional crossover problem in the 
weakly 
coupled chains.
In the final section, we give  concluding remarks.
Some technical details are left to Appendices.
\section{Dimensional Crossovers:   One-Particle vs. Two-Particle 
Crossovers}
The central problem here is how
the  weak interladder 
one-particle hopping, $t_{\perp}$, affects the low-energy asymptotics of 
the  
system. 
We switch on the intraladder interaction  and the interladder one-particle 
hopping, $t_{\perp}$,
 as perturbations to the 
system specified by the intraladder 
longitudinal (transverse) hopping, $t$($t'$)  (see Fig.~1).\cite{comment1}
Then two kinds of interladder processes, a one-particle process and a two-particle process, occur.
  We illustrate these processes 
in Fig.~2.
The one-particle and two-particle processes correspond to the interladder 
hopping of the 
single-particle excitation and that of the two-particle
 (particle-particle or particle-hole pair) excitation, respectively.
As the temperature scale decreases,  dimensional crossovers are induced by 
a one-particle process {\it or} 
a two-particle process.\cite{BY,YS,VMY,BC2}
If the former dominates the latter, an interladder coherent band motion occurs.
Then  the character of the ladder in one-dimension is completely lost.
On the other hand, if the latter dominates the former, 
 interladder coherent propagation of  a pair of composite particles occurs
and    the system transits to a long-range-ordered phase in the corresponding channel.
In Fig.~2, we show the   two-particle hopping in the $d$-wave
 superconducting channel,  which  corresponds to the 
interladder   Josephson tunneling of   bipolarons. We will see this is 
the most dominant two-particle process.

The isolated doped ladder system has a spin excitation gap,  which  
corresponds to  
the binding energy of
the bipolaron. Thus  the amplitude of the  interladder tunneling of a bipolaron  is roughly estimated as 
$t_{\perp}^2/\Delta_{\sigma}$,
where $\Delta_{\sigma}$ is the characteristic energy scale of the spin 
gap. On the other hand,
the amplitude of the one-particle hopping is $t_{\perp}$.
Consequently, the competition  is determined by the ratio, 
$t_{\perp}/\Delta_{\sigma}$, which is strongly  affected by 
one-dimensional
quantum fluctuations in the system.

\section{Scaling Equations for Interladder  One-Particle and Two-Particle 
Hopping Amplitudes}
To study the competition between the one-particle and two-particle 
crossovers,  we set up   scaling 
equations for the interladder one-particle  and two-particle 
hopping amplitudes.

\subsection{Action}
We start with the path integral representation of the partition function 
of the system,
$
Z=\ds\int{\cal D} e^{S},
$
where   the action  consists of four parts,
\beq
S=S^{(1)}_{\para}+S^{(2)}_{\para}+S^{(1)}_{\perp}+S^{(2)}_{\perp},\label{eqn:action}
\eeq
with 
$S^{(1)}_{\para}$, $S^{(2)}_{\para}$, $S^{(1)}_{\perp}$ and 
$S^{(2)}_{\perp}$ being the actions 
for the intraladder one-particle hopping, intraladder two-particle 
scattering,  interladder 
one-particle and interladder two-particle hoppings, respectively. 
 $\cal D$  symbolizes the measure of the path integral over the fermionic 
Grassmann variables.
 
The intraladder one-particle process is
 diagonalized with respect to the bonding ($B$) and antibonding ($A$) bands.
As shown in Fig.~3(a), we linearize the dispersion 
along the legs on the 
bonding and antibonding Fermi points, $\pm k_{Fm}$($m=A,B$). In Fig.~3,  
$R$ and $L$ 
denote the right-moving and left-moving branches, respectively.
The actions for the intra- and interladder one-particle hopping processes 
are written as 
\begin{eqnarray}
S^{(1)}_{\para}=\sum_{K} 
\sum_{m=A,B}\sum_{\sigma}\left[{\cal G}_{Lm}^{-1}(K_{\para})
L^{\ast}_{m\sigma}(K)L_{m\sigma}(K)\right.
+\left.{\cal G}_{Rm}^{-1}(K_{\para})
R^{\ast}_{m\sigma}(K)R_{m\sigma}(K)
\right],
\end{eqnarray}
and
\begin{eqnarray}
S^{(1)}_{\perp}=-\sum_{K}  \sum_{m=A,B}\sum_{\sigma}
\varepsilon_{\perp m}(\kperp)\left[
L^{\ast}_{m\sigma}(K)L_{m\sigma}(K)\right.
+\left.R^{\ast}_{m\sigma}(K)R_{m\sigma}(K)\right],
\end{eqnarray}
where  $R_{m\sigma}(K)$ and $L_{m\sigma}(K)$ are the Grassmann variables 
representing  the right- and left-moving electrons on
the  band $m$. The dispersions in the interladder action  are given by
$\varepsilon_{\perp A}(\kperp)=+t_{\perp}\cos\kperp$ and
$\varepsilon_{\perp B}(\kperp)=-t_{\perp}\cos\kperp$.
We use the notation
$K=(k_{\para},k_{\perp},\ii\varepsilon_{n})$
and
$K_{\para}=(k_{\para},\ii\varepsilon_{n})$
where $\varepsilon_{n}=(2n+1)\pi/\beta$ is a fermion  thermal frequency. 
We denote the momenta along 
the leg and the rung by $k_{\para}$ and $k_{\perp}$,  respectively. 
The intraladder Green functions are   written as
\begin{equation}
{\cal G}_{\nu m}(K_{\para})=[\ii\varepsilon_{n}-\varepsilon_{\nu 
m}(\kpara)]^{-1},
\end{equation}
where the linearized dispersions are given by  
$\varepsilon_{Rm}(\kpara)=v_{F}(\kpara-k_{Fm})$  and
$\varepsilon_{Lm}(\kpara)=v_{F}(-\kpara-k_{Fm})$. 
The Fermi momenta of the bonding and antibonding bands  
are given by $k_{FB}=k_{F}+t'/v_{F}$ and $k_{FA}=k_{F}-t'/v_{F}$, 
respectively. 
The Fermi velocities  in principle depend on the band index as
$v_{Fm}=2t\sin k_{Fm}$,
but we  assume throughout this work that $v_{Fm}=v_{F}$ and drop the band 
index, since
 the  difference in the  Fermi velocities does not affect the asymptotic 
 nature of the SGM phase at least for small $t'/t$.\cite{MF,BF}
 In all  the four branches($LB,LA,RA,RB$) of  linearized bands,  the 
energy variables, $\varepsilon_{\nu m}$($\nu=R,L;m=A,B$), run over the 
region,
$-E/2<\varepsilon_{\nu m}<E/2$, with
$E$ denoting the {\it bandwidth cutoff}.

The intraladder Hubbard repulsion generates  scattering processes 
depicted in Fig.~3(b).
The action for the intraladder two-particle scattering processes   is 
written as
\begin{eqnarray}
S^{(2)}_{\para}&=&
{2\pi v_{F}\over\beta } \sum  
g_{0}^{\sigma_{1}\sigma_{2}\sigma_{3}\sigma_{4}}L^{\ast}_{m\sigma_{1}}R^{\ast}_{m\sigma_{2}}R_{m\sigma_{3}} 
L_{m\sigma_{4}} 
+ {2\pi v_{F}\over\beta } \sum 
g_{f}^{\sigma_{1}\sigma_{2}\sigma_{3}\sigma_{4}} 
L^{\ast}_{m\sigma_{1}}R^{\ast}_{\bar m\sigma_{2}}R_{\bar m\sigma_{3}} 
L_{m\sigma_{4}}\non\\
&+&{2\pi v_{F}\over\beta } \sum 
g_{t}^{\sigma_{1}\sigma_{2}\sigma_{3}\sigma_{4}} 
L^{\ast}_{m\sigma_{1}}R^{\ast}_{m\sigma_{2}}R_{\bar m\sigma_{3}} L_{\bar 
m\sigma_{4}},
\end{eqnarray}
where 
 the summations are taken over band indices, spins, intraladder momenta 
and frequencies as
\begin{eqnarray*}
&&\sum 
L^{\ast}_{m_{1}\sigma_{1}}R^{\ast}_{m_{2}\sigma_{2}}R_{m_{3}\sigma_{3}} 
L_{m_{4}\sigma_{4}}\\
&=&
\sum_{K_{1},\cdots,K_{4}}\sum_{m_{1},\cdots,m_{4}}\sum_{\sigma_{1},\cdots,\sigma_{4}}\delta(K_{1}+K_{2}-K_{3}-K_{4})
L^{\ast}_{m_{1}\sigma_{1}}(K_{1})R^{\ast}_{m_{2}\sigma_{2}}(K_{2}) 
R_{m_{3}\sigma_{3}}(K_{3}) L_{m_{4}\sigma_{4}}(K_{4})
\end{eqnarray*}
with $m$ and $\bar m$ being different band indices.
The dimensionless intraladder scattering  strengths are given by
\beq
g^{\sigma_{1}\sigma_{2}\sigma_{3}\sigma_{4}}_{\mu}=
g^{(1)}_{\mu}\delta_{\sigma_{1}\sigma_{3}}\delta_{\sigma_{2}\sigma_{4}}
-g^{(2)}_{\mu}\delta_{\sigma_{1}\sigma_{4}}\delta_{\sigma_{2}\sigma_{3}},
\eeq
where dimensionless quantities 
$g^{(1)}_{\mu}$ and  $g^{(2)}_{\mu}$  denote backward and forward 
scattering strengths, respectively,
with the flavor indices,\cite{MF} $\mu=0,f,t$.
The usual scattering strengths with 
 dimension of the interaction energy are $2\pi v_{F}g_{\mu}^{(i)}$.
We neglect the  interband backward scattering terms like  
$
\sum 
g_{b}^{\sigma_{1}\sigma_{2}\sigma_{3}\sigma_{4}}L^{\ast}_{B\sigma_{1}}R^{\ast}_{A\sigma_{2}}R_{B\sigma_{3}} 
L_{A\sigma_{4}}
$
on the ground that  
these processes  do not seriously modify  the asymptotic nature of the 
C1S0 phase for finite $t'$\cite{MF,BF}. 
Since we neglect both of  the band index dependence of the Fermi 
velocities and the interband backward processes,
the intraladder transverse hopping, $t'$,  never appears explicitly in the 
present work.

The action for the interladder two-particle hopping processes are  
decomposed into CDW$\mu$, SDW$\mu$, SS$\mu$ (singlet superconducting) and 
TS$\mu$ (triplet superconducting) 
channels ($\mu=0,f,t$ specify the corresponding flavor indices) as
\begin{eqnarray}
S^{(2)}_{\perp}=
&-&{\pi v_{F}\over 2 }\sum_{\rm DW=CDW,SDW}\sum
\left[
{ V}^{\rm DW}_{0}{\cal O}^{mm\ast}_{\rm DW}{\cal O}^{mm}_{\rm DW}\right.
\left.
+{ V}^{\rm DW}_{t}{\cal O}^{m\bar m\ast}_{\rm DW}{\cal O}^{\bar m m}_{\rm 
DW}
+{ V}^{\rm DW}_{f}{\cal O}^{m \bar m\ast}_{\rm DW}{\cal O}^{m \bar m}_{\rm 
DW}\right]\non\\ 
&-&{\pi v_{F}\over 2 }\sum_{\rm S=SS,TS}\sum\left[{ V}^{\rm 
S}_{0}{\cal O}^{mm\ast}_{\rm S}{\cal O}^{mm}_{\rm S} 
+{ V}^{\rm S}_{t}{\cal O}^{mm\ast}_{\rm S}{\cal O}^{\bar m \bar m}_{\rm S} 
\right.
+\left.{ V}^{\rm S}_{f}{\cal O}^{\bar m m\ast}_{\rm S}{\cal O}^{\bar m 
m}_{\rm S}  
\right],
\label{eqn:TPH}
\end{eqnarray}
where ${ V}^{M}_{\mu}$ denotes the interladder pair tunneling amplitude 
for   channel $M\mu$.
$m$ and $\bar m$ denote different band indices. The summations are taken 
over band indices,  intra- and inter-ladder momenta and boson thermal frequencies, 
$Q=(q_{\para},q_{\perp},\ii \omega_{l})$, with $\omega_{l}=2l\pi /\beta$,  
\begin{eqnarray*}
\sum { V}^{M}_{\mu} {\cal O}^{m_{1}m_{2}\ast}_{M}{\cal O}^{m_{3}m_{4}}_{M}
=
\sum_{Q}\sum_{m_{1},\cdots,m_{4}}{ V}^{M}_{\mu} {\cal 
O}^{m_{1}m_{2}\ast}_{M}(Q)
{\cal O}^{m_{3}m_{4}}_{M}(Q).
\end{eqnarray*}
The corresponding composite field variables  are defined by
\beq
{\cal O}^{mm'}_{\rm CDW}(Q)&=&
\beta^{-1/2}\sum
R^{*}_{m\sigma}(K+Q)L_{m'\sigma}(K),\non\\
\vec{\cal O}^{mm'}_{\rm SDW}(Q)&=&\beta^{-1/2}\sum
R^{*}_{m\sigma}(K+Q){\vec\sigma}_{\sigma\sigma'}L_{m'\sigma'}(K),\\
{\cal O}^{mm'}_{\rm SS}(Q)&=&\beta^{-1/2}\sum
\sigma R_{m\sigma}(-K+Q)L_{m'\bar\sigma}(K),\non\\
\vec{\cal O}^{mm'}_{\rm TS}(Q)&=&\beta^{-1/2}\sum
\sigma 
R_{m\sigma}(-K+Q){\vec\sigma}_{\sigma\sigma'}L_{m'\bar\sigma'}(K),\non
\eeq
where   $\vec \sigma$ are Pauli matrices, $\bar\sigma=-\sigma$, and 
the summations are taken over spins and $K$.   

\subsection{Scaling of the intraladder two-particle scattering processes}
The idea of scaling is to eliminate  the short-wavelength degrees of 
freedom  to relate   
 effective actions  at successive energy   scales.
Based on the bandwidth cutoff regularization scheme, 
we parametrize the cutoff as $E(l)=E_{0}e^{-l}$ with the scaling 
parameter, $l$,
and study how  the  action is renormalized as $l$ goes from zero to 
infinity. 
Details on the derivation of the scaling equations are left to the 
Appendices. 

 Scaling equations for the intraladder scattering vertices, 
$g_{\mu}^{(i)}$, are generally written as\cite{Solyom}
\begin{eqnarray}
{d \ln g^{(i)}_{\mu}(l) \over dl}=z^{(i)}_{\para\mu}(l)-2z_{\para}(l).
\label{eqn:generalformforg}
\end{eqnarray}
Within the 3rd order PRG scheme, the first term on the r.h.s, 
$z^{(i)}_{\para\mu}(l)$, comes from the vertex correction diagrams 
represented in Figs.~4(a) and 4(b) for $i=1$ and 2, respectively.
The field rescaling procedure (see Appendix A) requires the second term, 
$2z_{\para}$, 
to be
\begin{eqnarray}  
z_{\para}(l)&=&
{{ g}_{0}^{(1)}}(l)^2
    +{{ g}_{0}^{(2)}}(l)^2
    +{{ g}_{f}^{(1)}}(l)^2
    +{{ g}_{f}^{(2)}}(l)^2
    +{{ g}_{t}^{(1)}}(l)^2
    +{{ g}_{t}^{(2)}}(l)^2\non\\
&&  -{ g}_{0}^{(1)}(l)
     { g}_{0}^{(2)}(l)
    -{ g}_{f}^{(1)}(l)
     { g}_{f}^{(2)}(l)
    -{ g}_{t}^{(1)}(l)
     { g}_{t}^{(2)}(l),\label{eqn:zpara}
\end{eqnarray}
which comes from the intraladder self-energy diagrams represented in 
Fig.~4(c).
We give a full detail on the derivation of $z_{\para}$ in Appendix B.
The full expressions for the scaling equations, 
(\ref{eqn:generalformforg}),  are given by
\begin{eqnarray}
&&{d\over dl}  g^{(1)}_{0}(l)
= -2( {  g^{(1)}_{0}}^{2}+  g^{(1)}_{t}  g^{(2)}_{t})+2(  g_{f}^{(1)}{ 
g_{t}^{(2)}}^2
-  g_{f}^{(1)}  g_{t}^{(1)}  g_{t}^{(2)})\non\\
&&\,\,\,\,\,\,\,\,\,\,\,\,\,\,\,\,\,\,\,\,\,\,\,\,\,\,\,\,\,\,\,-2{  
g}^{(1)}_{0}({{  g}_{0}^{(1)}}^2+{{  g}_{f}^{(1)}}^2
+{{  g}_{t}^{(1)}}^2+{{  g}_{t}^{(2)}}^2-{{  g}_{t}^{(1)}}{{  
g}_{t}^{(2)}}),\label{eqn:g01}\\
&&{d \over dl} g^{(2)}_{0}(l)=-(  {g^{(1)}_{0}}^{2}+  {g^{(1)}_{t}}^{2}+  
{g^{(2)}_{t}}^{2})
+2(  g_{f}^{(2)}{ g_{t}^{(2)}}^2
-  g_{t}^{(1)}  g_{t}^{(2)}  g_{f}^{(2)}) \non\\
&&\,\,\,\,\,\,\,\,\,\,\,\,\,\,\,\,\,\,\,\,\,\,\,\,\,\,\,\,\,\,\,
+2{  g_{t}^{(1)}}^2{  g_{f}^{(2)}}
-{  g_{0}^{(1)}}^3-{  g_{f}^{(1)}}^2{  g_{0}^{(1)}}
-{  g_{t}^{(1)}}^2{  g_{f}^{(1)}} 
-2{  g}^{(2)}_{0}({{  g}_{t}^{(1)}}^2+{{  g}_{t}^{(2)}}^2-{{  
g}_{t}^{(1)}}{{  g}_{t}^{(2)}}),\\
&&{d  \over dl}g^{(1)}_{f}(l) =
-2( { g^{(1)}_{f}}^{2}+{ g^{(1)}_{t}}^{2}-  g^{(1)}_{t}  g^{(2)}_{t})\non
\\&&\,\,\,\,\,\,\,\,\,\,\,\,\,\,\,\,\,\,\,\,\,\,\,\,\,\,\,\,\,\,
+2(  g_{0}^{(1)}{  g_{t}^{(2)}}^2
-g_{0}^{(1)}  g_{t}^{(1)}  g_{t}^{(2)} )
-2{  g}^{(1)}_{f}({{  g}_{0}^{(1)}}^2 +{{  g}_{f}^{(1)}}^2 
+{{  g}_{t}^{(1)}}^2+{{  g}_{t}^{(2)}}^2
-{{  g}_{t}^{(1)}}{{  g}_{t}^{(2)}}),\\
&&{d  \over dl} g^{(2)}_{f}(l)=-(  {g^{(1)}_{f}}^{2}-  {g^{(2)}_{t}}^{2})
+2(  g_{0}^{(2)}{  g_{t}^{(2)}}^2  
-  g_{t}^{(1)}  g_{t}^{(2)}  g_{0}^{(2)}  )\non \\
&&\,\,\,\,\,\,\,\,\,\,\,\,\,\,\,\,\,\,\,\,\,\,\,\,\,\,\,\,\,\,
+ 2{  g_{t}^{(1)}}^2{  g_{0}^{(2)}}
  -{  g_{t}^{(1)}}^2  g_{0}^{(1)}-{  g_{0}^{(1)}}^2  g_{f}^{(1)}
-{g_{f}^{(1)}}^3-2{g}^{(2)}_{f}({g_{t}^{(1)}}^2
+{{g}_{t}^{(2)}}^2-{{g}_{t}^{(1)}}{{g}_{t}^{(2)}}),\\
&&{d \over dl} g^{(1)}_{t}(l)=-2 (2 g^{(1)}_{t}g^{(1)}_{f} - 
g^{(1)}_{t}g^{(2)}_{f}-g^{(1)}_{f}g^{(2)}_{t} + 
g^{(1)}_{0}g^{(2)}_{t}+g^{(1)}_{t}g^{(2)}_{0})\non\\
&&\,\,\,\,\,\,\,\,\,\,\,\,\,\,\,\,\,\,\,\,\,\,\,\,\,\,\,\,\,\,
+2(2  g_{t}^{(1)}  g_{0}^{(2)}  g_{f}^{(2)}-  g_{0}^{(1)}  g_{t}^{(1)}  
g_{f}^{(2)}-g_{f}^{(1)}g_{t}^{(1)}  g_{0}^{(2)}) \\
&&\,\,\,\,\,\,\,\,\,\,\,\,\,\,\,\,\,\,\,\,\,\,\,\,\,\,\,\,\,\,
-2{g}^{(1)}_{t}({g_{0}^{(1)}}^2+{{  g}_{0}^{(2)}}^2+{{g}_{f}^{(1)}}^2+{{  
g}_{f}^{(2)}}^2
+{{g}_{t}^{(1)}}^2+{{g}_{t}^{(2)}}^2
-{g_{0}^{(1)}}{{g}_{0}^{(2)}}-{{g}_{f}^{(1)}}{{g}_{f}^{(2)}}-{{g}_{t}^{(1)}}{{g}_{t}^{(2)}}),\non\\
&&{d  \over dl} g^{(2)}_{t}(l)=-2(  g^{(1)}_{0}  g^{(1)}_{t}+
  g^{(2)}_{0}  g^{(2)}_{t}-  g^{(2)}_{f}  g^{(2)}_{t}) \non\\ 
&&\,\,\,\,\,\,\,+2(
2  g_{0}^{(2)}  g_{f}^{(2)}  g_{t}^{(2)}
-  g_{f}^{(1)}  g_{0}^{(2)}  g_{t}^{(2)}
-  g_{0}^{(1)}  g_{f}^{(2)}  g_{t}^{(2)}+2  g_{0}^{(1)}  g_{f}^{(1)}  
g_{t}^{(2)}
-  g_{t}^{(1)}  g_{f}^{(1)}  g_{0}^{(1)})\non \\
&&\,\,\,\,\,\,\,-2{  g}^{(2)}_{t}({{  g}_{0}^{(1)}}^2+{{  g}_{0}^{(2)}}^2+
{{  g}_{f}^{(1)}}^2+{{  g}_{f}^{(2)}}^2+{{  
g}_{t}^{(1)}}^2+{{g}_{t}^{(2)}}^2
-{{  g}_{0}^{(1)}}{{  g}_{0}^{(2)}}-{{  g}_{f}^{(1)}}{{  
g}_{f}^{(2)}}-{{g}_{t}^{(1)}}{{  g}_{t}^{(2)}}),\label{eqn:gt2}
\end{eqnarray}
\normalsize
where  we omitted the argument $l$ in the scattering strengths on the r.h.s.
The same equations are obtained by setting $g_{b}^{(1)}=g_{b}^{(2)}=0$ in 
Eq.(A5) of Ref.[13].

Starting with the Hubbard type initial condition,
\beq
 g^{(i)}_{\mu}(0)=\tilde U\equiv {U/ 4\pi v_{F}}>0,
\eeq
the scaling equations lead to the {\it strong coupling} fixed point\cite{MF}
\be
\begin{array}{ccc}
g^{(1)\ast}_{0}=-1,&g^{(1)\ast}_{f}=0,&g^{(1)\ast}_{t}=1,\\
g^{(2)\ast}_{0}=-{3-2\tilde U\over 4},&g^{(2)\ast}_{f}={1+2\tilde U\over 
4}, & g^{(2)\ast}_{t}=1.
\end{array}\label{FP}
\en

To obtain a physical description of the fixed point, it is useful to 
analyze the isolated Hubbard ladder system by means of the bosonization 
technique.\cite{MF,FL,KR,BF,Schulz1,NLK,YS,DVK,NN,ST,NO}
We introduce the phase fields, $\phi_{\nu}(x)$ and $\theta_{\nu}(x)$, 
where $\phi_{\nu}(x)$ and ${1\over \pi}\partial_{x}\theta_{\nu}(x)$ are 
conjugate,
\begin{eqnarray}
\left[\phi_{\nu}(x),{1\over \pi}\partial_{y}\theta_{\mu}(y)\right]=\ii 
\delta_{\mu\nu}\delta(x-y).
\end{eqnarray}
The mode-indices, $\mu,\nu=\rho A,\,\rho B,\,\sigma A,\,\sigma B,\,$ 
represent the charge- and spin-modes on the bands $A$ or $B$, respectively.
The charge- and spin-modes are combined into the total-mode (0-mode) and 
the relative-mode ($\pi$-mode) as
\begin{eqnarray}
\phi_{\rho\pm}=(\phi_{\rho B}\pm\phi_{\rho A})/\sqrt{2},\,\,\,
\phi_{\sigma\pm}=(\phi_{\sigma B}\pm\phi_{\sigma A})/\sqrt{2},
\end{eqnarray}
where $+$ and $-$ denote the 0-mode and $\pi$-mode, respectively.
Then the {\lq\lq}non-interacting {\rq\rq} part [including the 
$g_{\mu}^{(i)}$ processes ($\mu=0,f$) for $i=1$ with parallel spins and 
those for $i=2$] of the Hamiltionian for the isolated Hubbard 
ladder system is written as\cite{MF,KR,BF}
\begin{eqnarray}
{\cal H}_{0}=\sum_{\nu=\rho\pm,\sigma\pm}\int{dx\over 
2\pi}\left[u_{\nu}K_{\nu}\left(\partial_{x}\theta_{\nu} \right)^2
+{u_{\nu}\over K_{\nu}}\left(\partial_{x}\phi_{\nu}\right)^2\right].
\end{eqnarray}
The velocity, $u_{\nu}$, and the stiffness, $K_{\nu}$, of each mode  are 
given by
\begin{eqnarray}
u_{\nu}=v_{F}\sqrt{1-g_{\nu}^2},\,\,\,
K_{\nu}=\sqrt{{1+g_{\nu}\over 1-g_{\nu}}},
\end{eqnarray}
where
\begin{eqnarray}
g_{\rho\pm}=(g_{0}^{(1)}-2g_{0}^{(2)})\pm 
(g_{f}^{(1)}-2g_{f}^{(2)}),\,\,\,\,g_{\sigma\pm}=g_{0}^{(1)}\pm 
g_{f}^{(1)}.
\end{eqnarray}

The low-energy asymptotics of the isolated Hubbard ladder system is 
characterized by   
the strong coupling fixed point of three scattering strengths, $g^{(1)\ast}_{t}=$ 
$g^{(2)\ast}_{t}=$ $-g^{(1)\ast}_{0}=1$, which
correspond to
\begin{eqnarray}
1/K_{\rho-}=K_{\sigma+}=K_{\sigma-}=0.\label{FPKRS}
\end{eqnarray}
The scale-invariance of $K_{\rho+}=\sqrt{(1-2\tilde U)/( 1+2\tilde U)}$ 
means that the total-charge mode remains gapless.
On the other hand, $1/K_{\rho-}=0$ and $K_{\sigma+}=K_{\sigma-}=0$ mean 
that the phases $\theta_{\rho-}$ and $\phi_{\sigma\pm}$ are respectively 
locked, 
suggesting the corresponding modes acquire a gap.\cite{MF,KR} 
Therefore, in this phase,  only the total charge mode remains gapless and 
consequently 
the $d$-wave superconducting correlation  becomes the most dominant 
one\cite{MF,KR,BF,Schulz1} and 
the $4k_{F}$-CDW correlation becomes sub-dominant 
one\cite{KR,BF,Schulz1} at least when the intraladder correlation is weak. 
The phase characterized  by this fixed point is denoted by  the {\lq\lq}phase 
I{\rq\rq} by 
Fabrizio\cite{MF} and the {\lq\lq}C1S0 phase{\rq\rq} by Balents and 
Fisher.\cite{BF}
Throughout the present paper  we call this  phase  a spin gap metal (SGM) 
phase.
Exponents of the $d$-wave superconducting  
 and the $4k_{F}$-CDW correlations, $K_{S}$ and $K_{C}$, in the SGM 
phase 
 satisfy the {\it duality relation}, $K_{S}\cdot K_{C}=1$, suggesting the 
low-energy dynamics in the SGM phase is described with   {\it bipolarons},
each of which lies 
along the rung of 
the ladder.\cite{NN,TTR1}

\subsection{Scaling of the interladder one-particle hopping amplitude}
The  scaling of the interladder one-particle process is fully determined 
by the intraladder
 self-energy effects.  
The scaling equation, represented in Fig.~5, is written as
\begin{eqnarray}
{d\ln t_{\perp}(l)\over dl}=1-z_{\para}(l).\label{eqn:RGfort}
\end{eqnarray}
We solve the equation with the initial condition, 
\begin{equation}
 t_{\perp}(0)/E_{0}=\tilde t_{\perp0}.
\end{equation}
The scaling equation, (\ref{eqn:RGfort}), and the fixed point,  
(\ref{FP}), lead to
\beq
{d\ln t_{\perp}(l)\over dl}\stackrel{l\to\infty}{\longrightarrow}-\tilde 
U^{2}/2-{7/8},\label{eqn:DLcase}
\eeq
and consequently 
$t_{\perp}(l)$ {\it becomes  always irrelevant at the final stage of the 
scaling procedure}.
However,  at an early stage of scaling, the r.h.s of eq.(\ref{eqn:RGfort}) 
remains positive since the 
intraladder couplings do not  grow sufficiently as yet and consequently  
$t_{\perp}(l)$ grows. 
The r.h.s of (\ref{eqn:RGfort}) changes its sign from positive to negative 
in the course of renormalization and
then $\tilde t_{\perp}(l)$ begins to decrease and is finally scaled to 
zero.

If, in the course of the scaling, $\tilde t_{\perp}(l)\equiv 
t_{\perp}(l)/E_{0}$  attains an order of  unity around some crossover 
value of the scaling 
parameter, $l_{\rm cross}$, 
qualitatively specified by
\beq
\tilde t_{\perp}(l_{\rm cross})=1,\label{eqn:defoflc1}
\eeq 
the weakly coupled ladder picture   breaks down. 
Then, the notion of {\lq\lq}relevance{\rq\rq} or 
{\lq\lq}irrelevance{\rq\rq} of the interladder one-particle hopping
loses its literal meaning, since the PRG treatment for 
$t_{\perp}$ becomes inapplicable  for $l>l_{\rm cross}$.
If $l_{\rm cross}$ precedes the scaling parameter specifying the 
two-particle crossover,  
the interladder one-particle hopping becomes {\it coherent} and 
the system is regarded as scaled to a "two-dimensional"  system via the 
one-particle crossover.\cite{CB,BC1,BC2}

In   Fig.~6(a)   we illustrate the   scaling flows of 
$\tilde t_{\perp}(l)$  with a fixed value of the intraladder repulsion, 
$\tilde U=0.3$, and 
$\tilde t_{\perp0}=0.01,0.02,0.03,0.04$.
The peak positions are independent of $\tilde t_{\perp 0}$ and the peak 
heights are proportional to  $\tilde t_{\perp 0}$.
In   Fig.~6(b)   we illustrate the   scaling flows of 
$\tilde t_{\perp}(l)$  with a fixed  initial value of the 
interladder  one-particle hopping  
$\tilde t_{\perp 0}=0.01$ and $\tilde U=0.1,0.2,0.3,0.4,0.5$.
The peak positions and heights are sensitive to $\tilde U$.
We see from Fig.6 that for  small $\tilde U$, $\tilde t_{\perp}(l)$  
reaches unity even for small $\tilde t_{\perp0}$,
while for larger $\tilde U$, $\tilde t_{\perp}(l)$ are strongly suppressed 
and very large $t_{\perp}$ is required 
for $\tilde t_{\perp}(l)$ to attain an order of  unity.
As the intraladder correlations become stronger, the one-particle process 
is more severely suppressed.

\subsection{Scaling of the interladder two-particle hopping amplitudes}
The scaling equation for the interladder two-particle hopping amplitudes, 
$V_{\mu}^{M}$, are generally written as
\begin{equation}
{d  V^{M}_{\mu}(l) \over dl}=f_{\mu}^{M}(l)+z^{M}_{\perp\mu}(l)V_{\mu}^{M}(l),
\label{eqn:scalingof V}
\end{equation}
where $f_{\mu}^{M}$  denote the generators of $V^{M}_{\mu}$ for   channels  $M\mu$ ($M=$ CDW, SDW, SS, TS and 
$\mu=0,\,f,\,t$).
Note that $z^{M}_{\perp\mu}$ has $[V^{M}_{\mu}]^n$ contributions with $n=0$ 
and $n=1$. 
In Figs.7(a) and 7(b), we show the diagrammatic representation of 
$f_{\mu}^{M}$ for $M=$ CDW/SDW and SS/TS, respectively.
In Appendix C, we give the details on the derivation of $f_{\mu}^{M}$.

The leading-order scaling equations for $V_{\mu}^{M}$ are diagrammatically 
given in Figs.~8(a) and 8(b), for $M=$CDW/SDW and SS/TS, respectively.
Full expressions for the scaling equations, (\ref{eqn:scalingof V}), are 
written as
\begin{eqnarray}
{d { V}^{\rm DW}_{0}(l)\over dl}&=&
{1\over 4}\left[{\tilde t_{\perp}(l)g_{0}^{\rm DW}(l)}\right]^{2} \cos 
q_{\perp}
+{g_{0}^{\rm DW}}(l){ V}^{\rm DW}_{0}(l)
-{1\over 2}  {{ V}^{\rm DW}_{0}}(l)^{2},\label{eqn:CDW1} \\
{d { V}^{\rm DW}_{f}(l)\over dl}&=&
 - {1\over 4}\tilde t_{\perp}(l)^2\left[{g_{t}^{\rm DW}}(l)^2+ {g_{f}^{\rm 
DW}} (l)^2\right]
\cos q_{\perp}\\
&+& g_{t}^{\rm DW}(l){ V}^{\rm DW}_{t}(l)+ g_{f}^{\rm DW}(l) { V}^{\rm 
DW}_{f}(l)
-{1\over 2}\left[ {{ V}^{\rm DW}_{t}}(l)^{2}+ {{ V}^{\rm DW}_{f}} 
(l)^{2}\non
\right], \label{eqn:CDW2}\\
{d { V}^{\rm DW}_{t}(l)\over dl}&=&
 -{1\over 2}\tilde t_{\perp}(l) g_{t}^{\rm DW} (l)g_{f}^{\rm DW}(l)\cos 
q_{\perp}\\
&&+ g_{t}^{\rm DW}(l){ V}^{\rm DW}_{f}(l)+ g_{f}^{\rm DW} (l){ V}^{\rm 
DW}_{t} (l)
- { V}^{\rm DW}_{t}(l){ V}^{\rm DW}_{f}(l),\non \label{eqn:CDW3}\\
{d { V}^{\rm S}_{0}(l)\over dl}&=&-{1\over 4}\tilde t_{\perp}(l)^2
\left[{g_{0}^{\rm S}}(l)^{2}+{g_{t}^{\rm S}}(l)^{2}\right]\cos 
q_{\perp}
+ g_{0}^{\rm S}(l){ V}^{\rm S}_{0}(l)+ g_{t}^{\rm S} (l){ V}^{\rm 
S}_{t}(l) \\
&&-{1\over 2}\left[ {{ V}^{\rm S}_{0}}(l)^{2}+ { { V}^{\rm S}_{t}} (l)^{2}
\right],\non\\
{d { V}^{\rm S}_{f}(l)\over dl}&=&
{1\over 4}\tilde t_{\perp}(l)^2  g_{f}^{\rm S}(l)^{2}  \cos q_{\perp}
+ g_{f}^{\rm S}(l) { V}^{\rm S}_{f}(l) -{1\over 2} {{ V}^{\rm 
S}_{f}}(l)^{2} ,  \\
{d { V}^{\rm S}_{t}(l)\over dl}&=&
-{1\over 2}\tilde t_{\perp}(l)^2 g_{0}^{\rm S}(l) g_{t}^{\rm S}(l)\cos 
q_{\perp}
+ g_{0}^{\rm S}(l) { V}^{\rm S}_{t}(l)+ g_{t}^{\rm S} { V}^{\rm S}_{0}(l)
- { V}^{\rm S}_{0}(l){ V}^{\rm S}_{t}(l), \label{eqn:SST}
\end{eqnarray}
with DW$=$CDW/SDW and S$=$SS/TS.
Coupling strengths of the composite particles in  these channels are given by
\begin{eqnarray}
\left\{\begin{array}{c}
g^{\rm CDW}_{\mu}(l)=g^{(2)}_{\mu}(l)-2g^{(1)}_{\mu}(l),\,\,\,\,\,\,
g^{\rm SDW}_{\mu}(l)= g^{(2)}_{\mu}(l),\\
g^{\rm SS}_{\mu} (l)= -g^{(1)}_{\mu}(l)-g^{(2)}_{\mu}(l), \,\,\,\,\,\,
g^{\rm TS}_{\mu} (l)= g^{(1)}_{\mu}(l)-g^{(2)}_{\mu}(l). 
\end{array}\right.\label{eqn:cp}
\end{eqnarray}

We here in\-tro\-duce the com\-pos\-ite fields of the $s$\elvrm - and 
$d$\elvrm -wave su\-per\-con\-duct\-ing chan\-nels 
by\cite{MF,KR,BF,Schulz1,DVK,NN,NO}
\be
{\cal O}_{{\rm SC}s}={\cal O}_{\rm SS}^{BB}+{\cal O}_{\rm 
SS}^{AA},\,\,\,\,\,\,
{\cal O}_{{\rm SC}d}={\cal O}_{\rm SS}^{BB}-{\cal O}_{\rm SS}^{AA}.
\en
Cooper pairs in the $d$-wave channel (=bipolaron on the rung) are formed on the adjacent legs on 
the same rung (see Fig.~2).
The corresponding two-particle hopping amplitudes  are constructed  as
\be
{ V}^{{\rm SC}s}(l)= {1\over 2}\left[{ V}^{{\rm SS}}_{0}(l)+{ V}^{{\rm 
SS}}_{t}(l)\right],\,\,\,\,\,\,
{ V}^{{\rm SC}d}(l)= {1\over 2}\left[{ V}^{{\rm SS}}_{0}(l)-{ V}^{{\rm SS}}_{t}(l)\right],
\en
that satisfy the scaling equations 
\begin{eqnarray}
{d{ V}^{{\rm SC}s}\over dl}&=&-\left[\tilde t_{\perp}(l)
{g^{{\rm SC}s}(l)}\right]^{2}\cos q_{\perp}+2g_{0}^{{\rm SC}s}(l){V}^{{\rm 
SS}s}(l)-{1\over 2} V^{{\rm SS}s}(l)^{2},\label{eqn:SSs}\\  
{d{ V}^{{\rm SC}d}\over dl}&=&-\left[\tilde t_{\perp}(l)
{g^{{\rm SC}d}}(l)\right]^{2} \cos q_{\perp}+2{g_{0}^{{\rm SC}d}}(l){ 
V}^{{\rm SC}d} (l)-{1\over 2} {{ V}^{{\rm SC}d}}(l)^{2}, \label{eqn:SSd}
\end{eqnarray}
with the coupling strengths 
\beq 
{g^{{\rm SC}s}}(l)={1\over 2}\left[g_{0}^{\rm SS}(l)+g_{t}^{\rm 
SS}(l)\right],\,\,\,\,\,\,
{g^{{\rm SC}d}}(l)={1\over 2}\left[g_{0}^{\rm SS}(l)-g_{t}^{\rm 
SS}(l)\right],
\eeq
which are scaled to the fixed point, 
\beq
{g^{{\rm SC}s\ast}}=-(1+2\tilde U)/8,\,\,\,\,\,\,
{g^{{\rm SC}d\ast}}=(15-2\tilde U)/ 8.\label{FPSCD}
\eeq

We have solved  simultaneously 
the scaling equations, (\ref{eqn:g01}) $\sim$ (\ref{eqn:gt2}), 
(\ref{eqn:RGfort}), (\ref{eqn:CDW1}) $\sim$ (\ref{eqn:SST}),
(\ref{eqn:SSs}) and (\ref{eqn:SSd})  with the initial conditions
\begin{eqnarray}
g_{\mu}^{(i)}(0)=U/4\pi v_{F}\equiv \tilde U,\,\,\,\,\,\,
t_{\perp}(0)/E_{0}=\tilde t_{\perp0},\,\,\,\,\,\,
V^{M}_{\mu}(0)=0.\label{eqn:IC}
\end{eqnarray}
We put $q_{\perp}=0$ and $q_{\perp}=\pi$ for the superconducting and 
density-wave channels, respectively.
In Figs.9(a) $\sim$ (c), we show the scaling flows of the coupling  strengths for the 
composite fields, (\ref{eqn:cp}),
and $V^{M}_{\mu}(l)$ for  $\tilde U=0.2, 0.3, 0.4$ and 
$\tilde t_{\perp0}=0.01$,
where  the vertical broken line corresponds to the scaling parameter at 
which $V^{{\rm SC}d}$ diverges.
We clearly see that  the divergence of $V^{{\rm SC}d}$ always occurs at 
the highest energy scale. 
This situation is quite reasonable on physical
grounds that the interladder pair tunneling stabilizes the most dominant 
intraladder correlation, i.e, the $d$-wave superconducting correlation.
As is seen from Figs.~9(d) $\sim$ (f), the competition among the   channels 
other than the SC$d$ channel is rather subtle.
Below we  focus on  the $d$-wave superconducting channel.

The first, second and the third terms on the r.h.s of eq.(\ref{eqn:SSd})
play  separate roles in the early, intermediate and final stages of the 
scaling of $V^{{\rm SC}d}$.
At the early  stage,   the first term generates a finite magnitude of 
$V^{{\rm SC}d}$. 
At the intermediate stage, the second term induces an exponential growth 
of $V^{{\rm SC}d}$.
At the final stage, the third term  causes divergence of  ${ V}^{{\rm SCd}}$ at a 
critical scaling parameter $l_{c}$ defined  by
\beq
{ V}^{{\rm SC}d}(l_{c})=-\infty\label{eqn:defoflc2}.
\eeq
\section{Phase Diagram of Weakly Coupled Hubbard Ladders}
Based on the scaling flows of   $\tilde t_{\perp}(l)$ and $V^{{\rm SC}d}$,
we show the competition between the one- and two-particle crossovers.
The scaling flows of the intraladder system toward the SGM phase are best 
visualized through the scaling
flow of $K_{\sigma+}$, the stiffness of the total-spin mode.
We here again note that $K_{\sigma+}=0$ corresponds to the fully developed 
spin gap at the low-energy
limit.
In Fig.~10, we illustrate the scaling flows of $K_{\sigma+}$, $\tilde 
t_{\perp}(l)$ and 
$V^{{\rm SC}d}$ with various initial conditions. 
We see from Fig.~10 that, for $(\tilde U,\tilde t_{\perp0})$ =(0.3,0.01), 
(0.3,0.02), (0.4,0.01) and (0.5,0.01),
the two-particle crossover occurs at $l=l_{c}$.  
On the other hand, for $(\tilde U,\tilde t_{\perp0})$ =(0.3,0.03), 
(0.3,0.04), (0.1,0.01) and (0.2,0.01),
$\tilde t_{\perp}(l)$ exceeds unity at $l_{\rm cross}$ before $V^{{\rm 
SC}d}$ diverges.
Then the one-particle crossover occurs.

The scaling parameter is identified with the absolute temperature   as
$
l=\ln {E_{0}\over T}.
$
Thus  we   define 
the one-particle {\it crossover temperature}, $T_{\rm cross}$, and the 
$d$-wave superconducting 
{\it transition temperature}, $T_{c}$, as
\beq
\begin{array}{c}
T_{\rm cross}=E_{0}e^{-l_{\rm cross}},\\
T_{c}=E_{0}e^{-l_{c}}.
\end{array}
\eeq 
We obtain  phase diagrams of the system with respect to $(\tilde U,\tilde 
t_{\perp0})$ and the reduced temperature,
$\tilde T=T/E_{0}$.

\subsection{$t_{\perp0}$-$T$ phase diagram}
First we show, in Fig.~11, the phase diagram  spanned by $\tilde 
t_{\perp0}$ and $\tilde T$ for   $\tilde U=0.3$.
Roughly speaking, we may regard increasing $\tilde t_{\perp0}$ as applying 
the pressure, under which the bulk superconductivity was
actually observed in a doped spin ladder, 
Sr$_{14-x}$Ca$_{x}$Cu$_{24}$O$_{41}$.\cite{AG,MOTT}
We found that  there exists a crossover value of the interladder 
one-particle hopping, $\tilde t_{\perp c}\sim 0.025$.

For $0<\tilde t_{\perp0}<\tilde t_{\perp c}$,  the phase transition into 
the $d$-wave superconducting (SCd) phase   occurs
 at a finite transition temperature, $T_{c}$,  via the condensation of
bipolarons driven by  the two-particle 
crossover.  
In the temperature region, $T<T_{c}$, interladder coherent   Josephson 
tunneling of the bipolarons occurs. 
Now we must use caution in  identifying  the finite temperature phase 
above $T_{c}$, where the system is in 
the {\it isolated ladder regime.}
As the temperature scale decreases,   the isolated ladder systems are {\it 
gradually} scaled 
to their low-energy asymptotics, the  SGM phase, toward the zero 
temperature.
The gradual change of   darkness in the SGM phase in Fig.11 
schematically illustrates this situation.
The SGM phase is characterized by the strong coupling fixed point of 
intraladder scattering strengths, $g_{t}^{(1)}=g_{t}^{(2)}=-g_{0}^{(1)}=1$,
or $1/K_{\rho-}=K_{\sigma+}=K_{\sigma-}=0 $  (see (\ref{FP}) and 
(\ref{FPKRS})).
We see from Fig.10 that the critical scaling parameter $l_{c}$ is always 
in the region around 
which $K_{\sigma+}$ almost reaches its fixed point value.
Thus we expect that the spin gap is well developed near  $T_{c}$ and
the phase transition   at $T_{c}$ can be identified with  the transition 
from 
the SGM phase to the SCd phase.
Within the framework of the PRG approach, however,
we cannot say for certain whether the spin gap survives in the SCd phase 
or not.

For $\tilde t_{\perp c}<\tilde t_{\perp0}$,   the system undergoes the crossover to the  2D phase
via the one-particle crossover. 
The temperature scale of the one-particle crossover is not low enough   to 
ensure a well-developed spin gap  since
the one-particle crossover takes place at an early stage of scaling, where 
the intraladder couplings are still far away from their strong coupling 
values, as
can be seen from Fig.10.
Thus it is disputable to  assign the phase above $T_{\rm cross}$ to the 
SGM phase.
In the temperature region, $T<T_{\rm cross}$ the transverse {\it coherent} 
band motion occurs. In section 4.4, we give brief comments on the physical nature of the 2D phase.

In this paper, we consider the case where $t'$ is much larger than $t_{\perp0}$ 
and is not so large as only one band to cut the Fermi surface (these conditions 
are actually satisfied in the real Sr$_{14-x}$Ca$_{x}$Cu$_{24}$O$_{41}$ 
compounds\cite{AT}). Then, the (hypothetical) isolated ladder system is always 
scaled to the SGM phase. When $t'$ decreases and becomes comparable to 
$t_{\perp0}$, the system may be regarded as a coupled chain system, since, 
in such a case, the alternation of the transverse hopping would be unimportant. 
As we shall discuss in \S 5, 
in the coupled chain system, the one-particle crossover converts the system 
to the 2D phase as far as $\tilde U$ is not extremely large. Then the SCd
region 
would shrink with increasing $t'$ to be connected with the case of comparable 
$t'$ and $t_{\perp0}$. However, our result (Fig.~11) would 
not be qualitatively changed.

\subsection{$U$-$T$ phase diagram}
In Fig.~12, we show the phase diagram spanned by $\tilde U$ and $\tilde T$
for   $\tilde 
t_{\perp0}=0.01$.
We see that there exists a crossover value of the intraladder repulsion, 
$\tilde U_{c}\sim 0.22$.
For $0<\tilde U<\tilde U_{c}$, the system undergoes the crossover to the 2D  phase   
via the one-particle crossover,
while for $\tilde U_{c}<\tilde U$, the SGM phase
transits 
to the  SCd phase via the two-particle crossover.

\subsection{Crossover values of the interladder one-particle hopping: 
$\tilde t_{\perp c}$}
In Fig.~13, we show how the crossover value, $\tilde t_{\perp c}$, 
introduced in Fig.~11,
depends on $\tilde U$.
For $\tilde t_{\perp0}>\tilde t_{\perp c}$, the system undergoes the crossover to the 
2D phase, while
for  $\tilde t_{\perp0}<\tilde t_{\perp c}$, the SGM phase transits to the 
SCd phase.
$\tilde t_{\perp c}$ is always finite for $\tilde U>0$ and becomes zero 
only for $\tilde U=0$.
Thus we see that {\it the crossover value, $\tilde t_{\perp c}$, always exists for 
$\tilde U>0$.}
The SCd phase vanishes only for $\tilde U=0$, where the scaling equation, 
(\ref{eqn:RGfort}), gives  
$\tilde T_{\rm cross}=\tilde t_{\perp0}$.

\subsection{Remarks on the 2D phase} 
Here we give a few remarks on the 2D phase.
In the 2D phase the physical properties of the 
system would strongly depend on the shape of the 2D Fermi surface which 
may be characterized by
the misfit of the rungs in the neighboring ladders  which actually exists  
in real Sr$_{14-x}$Ca$_{x}$Cu$_{24}$O$_{41}$ compounds (see Fig.~14(a)).
The band structure of the system is characterized by bonding (B) and antibonding (A) dispersions,
\begin{eqnarray}
\varepsilon_{A,B}({\vec k})=-2t\cos k_{\para}\pm\sqrt{t'^2+4t''^2\cos^2 {k_{\perp}\over 2}
+2t't''\cos {k_{\perp}\over 2}\cos k_{\para}},
\end{eqnarray}
where the $+$ and $-$ signs are taken for $A$ and $B$ bands, respectively.
It was pointed out by Yamaji\cite{yamaji} that in the two-dimensional two-band system, a superconducting transition
is possible via the so-called Suhl-Kondo mechanism,\cite{suhl,kondo} where 
 the interband exchange-like interaction  is strongly enhanced by an {\it interband nesting}.
Then the pairing interaction is caused by  the processe expressed by the interband particle-hole ladder 
diagrams.
Quite recently it was  reported\cite{kon} 
that, by using a numerical fluctuation-exchange (FLEX) method,
a superconducting transition actually becomes possible in a coupled Hubbard ladder system with the same configuration
 as that shown in Fig.~14(a),
where the two-band structure of the system is taken into account.

To check how the interband polarization is enhanced by the interband nesting, we calculate the static
interband polarization function written as
\begin{eqnarray}
\chi_{AB}(q_{\para},q_{\perp})={1\over 2}\sum_{\vec k}{
\tanh[{\beta\over 2}(\varepsilon_{A}({\vec k}+{\vec q}/2)-\mu)]-
\tanh[{\beta\over 2}(\varepsilon_{B}({\vec k}-{\vec q}/2)-\mu)]
\over
\varepsilon_{A}({\vec k}+{\vec q}/2)-\varepsilon_{B}({\vec k}-{\vec q}/2)
},
\end{eqnarray}
In Fig.~14(b), we show $\chi_{AB}(q_{\para},q_{\perp})$     
on the 1st Brillouin zone along the line $\Gamma(0,0)\to {\rm X}(\pi,0) \to {\rm M}(\pi,\pi) \to \Gamma$
for for $t=t'$ and $t''/t=0,0.2,0.3$ with a chemical potential $\mu=0$ and a temperature, $\beta^{-1}=T=10^{-5}t$.
We see that an increasing $t''$ decreases the degree of interband nesting and consequently 
the interband polarization function decreases.
Then the exchange-like pairing  interaction would also decrease.
Thus,  when a superconducting transition via Suhl-Kondo mechanism  becomes possible in the 2D phase,
the supercondicting transition temperature, $T_{c}$, should decrease with an increasing interladder coupling.

\section{Comparison with Dimensional Crossovers in Weakly Coupled  Chains}
In this section we compare the dimensional crossovers in the present system
 with  those in the  weakly coupled Hubbard 
chains.
Although the PRG approach to the dimensional crossover problem 
for the weakly coupled chains
has   been extensively studied by Broubonnais and Caron,\cite{BC2}   it 
is   
instructive to re-examine their work, clarifying  the difference
in the nature of  the dimensional crossovers in the two cases.

The action  for the weakly coupled chain system consists of four parts,
\beq
S_{\rm chain}=S^{(1)}_{{\rm chain}\para}+S^{(2)}_{{\rm 
chain}\para}+S^{(1)}_{{\rm chain}\perp}+S^{(2)}_{{\rm 
chain}\perp},\label{eqn:caction}
\eeq
where 
$S^{(1)}_{{\rm chain}\para}$, $S^{(2)}_{{\rm chain}\para}$, $S^{(1)}_{{\rm 
chain}\perp}$ and 
$S^{(2)}_{{\rm chain}\perp}$ denote the actions
for the intrachain one-particle hopping, intrachain two-particle 
scattering,  interchain
one-particle and interchain two-particle hoppings, respectively.
As shown in Fig.~15(a), we linearize the dispersion along the chains on the 
two Fermi points, $\pm k_{F}$. 
The intrachain Hubbard repulsion generates  scattering processes 
depicted in Fig.~15(b).
The processes are specified by dimensionless quantities $g^{(1)}$ 
and  $g^{(2)}$  
denoting backward and forward scattering strengths, respectively.\cite{Solyom}
The usual scattering strengths with 
 dimension of the interaction energy are $\pi v_{F}g^{(i)}$.

The action for the interchain two-particle hopping processes are  
decomposed into CDW, SDW, SS(singlet superconducting) and TS(triplet 
superconducting) channels
and is written, instead of  (\ref{eqn:TPH}), as
\begin{eqnarray}
S^{(2)}_{{\rm chain}\perp}=
-{\pi v_{F}\over 4 }\sum_{\rm D=CDW,SDW}\sum_{Q}
{ V}^{\rm D}{\cal O}^{\ast}_{\rm D}{\cal O}_{\rm D} 
-{\pi v_{F}\over 4 }\sum_{\rm S=SS,TS}\sum_{Q}{ V}^{\rm S}{\cal 
O}^{\ast}_{\rm S}{\cal O}_{\rm S},
\label{eqn:cint}
\end{eqnarray}
where ${V}^{M}$ denotes the  amplitude of the interchain two-particle 
hopping   for   channel $M$.
The corresponding composite field variables  are defined by
\beq
{\cal O}_{\rm CDW}(Q)&=&
\beta^{-1/2}\sum
R^{*}_{\sigma}(K+Q)L_{\sigma}(K),\non\\
\vec{\cal O}_{\rm SDW}(Q)&=&\beta^{-1/2}\sum
R^{*}_{\sigma}(K+Q)\vec{\sigma}_{\sigma\sigma'}L_{\sigma'}(K),\\
{\cal O}_{\rm SS}(Q)&=&\beta^{-1/2}\sum
\sigma R_{\sigma}(-K+Q)L_{\bar\sigma}(K),\non\\
\vec{\cal O}_{\rm TS}(Q)&=&\beta^{-1/2}\sum
\sigma R_{\sigma}(-K+Q)\vec{\sigma}_{\sigma\sigma'}L_{\bar\sigma'}(K),\non
\eeq
where  $\vec \sigma$ are  Pauli matrices, $\bar\sigma=-\sigma$, and the summations 
are taken over spins and $K$ with
$K=(k_{\para},k_{\perp},\ii \varepsilon_{n})$ for each
$Q=(q_{\para},q_{\perp},\ii \omega_{l})$ with  
 fermion and boson thermal frequencies, $\varepsilon_{n}=(2n+1)\pi /\beta$ 
and $\omega_{l}=2l\pi /\beta$, respectively.
We denote the momenta along and perpendicular 
to the chains by $k_{\para}$ and $k_{\perp}$, respectively.

The  diagrams which give the scaling equations for the intrachain 
scattering processes are the same as in Fig.~4, but some of them are 
canceled out
 to give, instead of (\ref{eqn:g01}) $\sim$ (\ref{eqn:gt2}), simple 
scaling equations\cite{Solyom}
\begin{eqnarray}
{d g^{(1)}(l)\over dl}&=&-g^{(1)}(l)^2-{1\over 
2}g^{(1)}(l)^3,\label{eqn:g1TL}\\
{d g^{(2)}(l)\over dl}&=&-{1\over 2}g^{(1)}(l)^2-{1\over 
4}g^{(1)}(l)^3.\label{eqn:g2TL}
\end{eqnarray}
Starting with the Hubbard type initial condition
\beq
 g^{(i)}(0)=U/\pi v_{F}=4\tilde U>0,
\eeq
the scaling equations lead to the   fixed point
\be
g^{(1)\ast}=0,\,\, g^{(2)\ast}=2\tilde U.\label{TLFP}
\en
The fixed point characterizes the 
Tomonaga-Luttinger  liquid (TL) phase.
Since $g^{(1)}$ monotonically decreases, $g^{(1)3}$ terms in 
(\ref{eqn:g1TL}) and (\ref{eqn:g2TL})
are not essential for the scaling property so that we can drop them.
Then we obtain
\begin{eqnarray}
g^{(1)}(l)&=&{4\tilde U\over 1+4\tilde U l},\label{eqn:g1sol}\\
g^{(2)}(l)&=&2\tilde U+{2\tilde U\over 1+4\tilde U l}.\label{eqn:g2sol}
\end{eqnarray}

The TL phase is different from the SGM phase of the Hubbard ladder in that 
it belongs to the {\it weak-coupling universality class} and is {\it 
gapless}.
Reflecting the weak-coupling nature, the intrachain self-energy effects 
are much weaker than in the  ladder system.
Consequently the strong suppression of the interchain one-particle process 
which is characteristic of the   coupled ladders never occurs in the 
coupled chains.
Furthermore, since   the TL phase is  gapless, the interchain two-particle 
hopping amplitude is ill defined (see the discussion given in \S 2).

The scaling equation  for the interchain one-particle hopping 
amplitude, $t_{\rm chain\perp}$,\cite{BC2} is, instead of 
(\ref{eqn:RGfort}),
written as
\begin{eqnarray}
{d\ln t_{\rm chain\perp}(l)\over dl}
=1-z_{\rm chain\para}(l),\label{eqn:RGfortchain}
\end{eqnarray}
with
\begin{eqnarray}
z_{\rm chain\para}(l)={1\over 
4}\left[g^{(1)}(l)^2+g^{(2)}(l)^2-g^{(1)}(l)g^{(2)}(l)\right].
\end{eqnarray}
Using (\ref{eqn:g1sol}) and (\ref{eqn:g2sol}),  we obtain
\begin{equation}
\tilde t_{\rm chain\perp }(l)=\tilde t_{\rm chain\perp 0}\exp 
\left[(1-\tilde U^2)l +{3\tilde U\over 4}\left({1\over 1+4\tilde U 
l}-1\right)\right].
\end{equation}
Thus, contrary to   the  coupled-ladder case,  $t_{\perp}$ 
becomes {\it relevant}  for   weak
 repulsion, $\tilde U<1$, where the PRG scheme is reliable.
We show this situation in the upper half plane of Fig.16, where the 
scaling flows of $t_{\rm chain\perp}(l)$
are shown for different $\tilde U=0.1,\,0.2,\,0.3$ with  $\tilde t_{\rm 
chain\perp0}=0.01$.
The one-particle crossover temperature is defined by
$
\tilde T^{\rm chain}_{\rm cross}=e^{-l^{\rm chain}_{\rm cross}},
$
where
$l^{\rm chain}_{\rm cross}$ is determined by 
$\tilde t_{\rm chain\perp}(l^{\rm chain}_{\rm cross})=1$.\cite{BC2}
The scaling behavior of $\tilde t_{\rm chain\perp }(l)$ is also expressed 
in terms of the anomalous exponent, $\theta$, as
$\tilde t_{\rm chain\perp }(l)=\tilde t_{\rm 
chain\perp0}\exp[(1-\theta)l]$ which gives
$
T^{\rm chain}_{\rm cross}=E_{0}[t_{\rm chain\perp0}/E_{0}]^{1/(1-\theta)}.
$
The  exact solution\cite{Schulz2,KSY,FK} tells us $\theta$ satisfies 
$\theta\leq 1/8$, which
again indicates $t_{\perp}$ is always relevant.

Structures of the leading-order   scaling equations for $V^{M}$ ($M=$CDW, 
SDW, SS, TS) are the same as those shown in 
Fig.~8,  which give\cite{BC2}
\begin{eqnarray}
{d { V}^{\rm DW}(l)\over dl}={1\over 2}\left[\tilde t_{\rm chain 
\perp}(l)g^{\rm DW}(l)\right]^{2}\cos q_{\perp}
+{1\over 2}{g}^{\rm DW}(l){ V}^{\rm DW}(l)
-{1\over 4}  {{ V}^{\rm DW}}(l)^2,
\end{eqnarray}
for DW$=$CDW or SDW and
\begin{eqnarray}
{d { V}^{\rm S}(l)\over dl}=-{1\over 2}\left[\tilde t_{\rm chain 
\perp}(l)g^{\rm S}(l)\right]^{2}\cos q_{\perp}
+ {1\over 2}g^{\rm S} (l)V^{\rm S}(l) -{1\over 4}V^{\rm S} (l)^2,
\end{eqnarray}
for S$=$SS or TS. We put $q_{\perp}=0$ and $q_{\perp}=\pi$ for the 
superconducting and density-wave channels, respectively.
Coupling strengths of the corresponding composite particles   are given by
\begin{eqnarray}
\left\{\begin{array}{c}
g^{\rm CDW}(l)=g^{(2)}(l)-2g^{(1)}(l),\,\,\,\,\,\,
g^{\rm SDW}(l)= g^{(2)}(l),\\
g^{\rm SS} (l)= -(g^{(1)}(l)+g^{(2)}(l)), \,\,\,\,\,\,
g^{\rm TS}(l)= g^{(1)}(l)-g^{(2)}(l). 
\end{array}\right.
\end{eqnarray}
In the lower half plane of Fig.~16, we show the  scaling flows of 
$V^{\rm SDW}$. 

In Figs.17(a) and (b), we show the phase diagrams of the weakly coupled chains 
for   $\tilde U=0.3$ and for $\tilde t_{\rm chain 
\perp0}=0.01$, which
correspond to Figs.11 and 12 in   the weakly coupled ladders, 
respectively.
For  small $\tilde U$ and $\tilde t_{\rm chain\perp0}$,    where the PRG 
scheme is reliable,  the  system
 always undergoes a crossover to a two-dimensional 
 phase (2D phase) via the one-particle processes.
We show in Fig.17 for guidance the temperature, $\tilde T_{\rm SDW}$,
 at which $V^{\rm SDW}$ diverges.

For the reasons already stated in \S 4.1,   it needs  care to 
identify  the finite temperature phase
 above $T_{\rm cross}^{\rm chain}$.
We can identify it with the TL phase characterized by the fixed point (\ref{TLFP}) only for very small $\tilde t_{\rm chain\perp 0}$ or large $\tilde U$,
where the one-particle crossover occurs at a very low temperature scale.
For large $\tilde t_{\rm chain\perp 0}$ or small $\tilde U$,
 the one-particle crossover occurs at a very early stage of the scaling 
where the intrachain scattering processes are still far away from
their fixed point corresponding to the low-energy asymptotics of the system.
The ambiguity in regarding the phase at $T>T_{\rm cross}^{\rm chain}$ as the TL phase
 has some relevance to  well known objections against  the 
PRG approach to the present problem, as first emphasized by
Anderson.\cite{PWA}
He stressed that it is crucial to treat the intrachain interaction exactly 
{\it before} switching on the interchain one-particle hopping.
The PRG scheme misses this point, since  we there treat the interaction 
and the interchain one-particle hopping, {\it on equal footing}, as 
perturbations to the 
{\it   one-dimeniosnal free fermion system}. 
Along Anderson's claim, by using the exact propagator of the TL liquid 
and  calculating the momentum distribution function,  
Castellani {\it et al.}~\cite{CCM} demonstrated 
that the TL phases are unstable with respect to arbitrary small interchain 
one-particle hopping, even if the interaction is strong.
Their conclusion supports the results obtained through the PRG scheme, 
contraly to the expectation of Anderson.

On the other hand,  the effects of the interchain one-particle hopping  
have been studied in
view of {\lq\lq}coherence{\rq\rq} or {\lq\lq}incoherence{\rq\rq} 
rather than its {\lq\lq}relevance{\rq\rq} or {\lq\lq}irrelevance{\rq\rq}.\cite{CSA,AMT}
With the aid of the   knowledge of the two-level system, it was 
demonstrated\cite{CSA} 
that there exists a 
critical value of the inter-Luttinger-liquid hopping below which
there is no coherent one-particle hopping between the liquids, which 
contradicts to the
 PRG results.
At present, however, it remains an unsettled question whether the TL phase 
is stable against the interchain
one-particle hopping or not.

\section{Concluding Remarks}
In the present paper we applied the perturbative renormalization group 
(PRG) method to discuss dimensional crossovers in Hubbard ladders coupled 
via
weak interladder one-particle hopping, $t_{\perp}$.
We set up and solved the scaling equations for the interladder 
one-particle and two-particle hopping amplitudes by treating the 
intraladder interactions and
the interladder one-particle hopping as perturbations to the free electron 
system on the isolated 
ladders.
We found that {\it the scaling flow toward the spin gap metal phase (SGM 
phase) in the isolated Hubbard 
ladder strongly suppresses the interladder 
one-particle process and consequently,   for  any finite intraladder 
Hubbard repulsion, $U>0$, 
there   exits a finite crossover value of the interladder one-particle 
hopping,
$t_{\perp c}$.} 

For $0<t_{\perp}<t_{\perp c}$,    the bulk $d$-wave superconducting phase 
is 
stabilized via the condensation of  bipolarons, where  interladder Josephson
tunnleling of bipolarons occur.
The superconducting transition occurs at the low temperature scales where 
the intraladder scattering processes nearly 
attain their strong coupling values (see Fig.10), which suggests  the 
spin gap is already well developed around the temperature region.
For $t_{\perp c}<t_{\perp}$, the system undergoes the crossover from the spin gap 
metal phase to the two-dimensional phase (2D phase) where the interladder 
one-particle hopping becomes
coherent and one-dimensionality of the isolated ladder is  completely 
lost. 
The temperature scales of the one-particle crossover is, however,   not low
enough  to ensure a well-developed spin gap (see Fig.10), since
the one-particle crossover occurs at an early stage of scaling, where the 
intraladder system does not
 approach the strong coupling region as yet.
Thus it is disputable  whether we can assign the region $T>T_{\rm cross}$ 
to the SGM phase or not.

Our results are best summarized in the $t_{\perp}$-$T$ phase diagram, 
Fig.11. Roughly speaking, we may regard   increasing $t_{\perp}$ as applying   pressure.
In the doped spin lad\-der,
 Sr$_{2.5}$Ca$_{11.5}$Cu$_{24}$O$_{41}$,
the superconductivity sets in below 10 K under     3.5 GPa $\sim$ 8 GPa. 
The observed  transition temperature, $T_{c}^{\rm obs}$, increases with the pressure
and reaches the maximum 6K around the optimal pressure $P_{\rm opt}\sim$ 4.5 GPa.\cite{MOTT} 
Our result in Fig.11 qualitatively reproduces the increase of $T_{c}^{\rm obs}$ for $P<P_{\rm opt}$.
For $P>P_{\rm opt}$,   the $T_{c}^{\rm obs}$ gradually decreases and  
the resistivity along the ladder, $\rho_{c}$, shows gradual change from $T$-linear to $T^2$-dependence with increasing
the applied pressure.\cite{MOTT} This fact strongly suggests that for $P>P_{\rm opt}$ the system is regarded as
an anisotropic 2D Fermi liquid.  
In our scheme, the existence of $\tilde t_{\perp c}$ would correspond to the
existence of  $P_{\rm opt}$. As was suggested in section 4.4,  it is reasonable to suppose
that in the 2D phase a superconducting instability via Suhl-Kondo mechanism\cite{yamaji}   becomes possible due
to the interband exchange-like interaction which is strongly enhanced by an {\it interband nesting}.
 Then the applied pressure tends to
decrease the degree of the interband nesting and consequently the transition temperature should decrease with increasing
the applied pressure.

We also discussed difference between   dimensional crossovers in weakly 
coupled Hubbard chains and ladders within the framework of the PRG.
The difference  originates from different  universality classes to which 
the corresponding isolated systems belong.
The isolated Hubbard chain belongs to the weak coupling Tomonaga-Luttinger 
liquid phase which is gapless, while the isolated Hubbard ladder belongs 
to the
strong coupling phase with the spin gap.
In the former case, the two-particle process is always dominated by the 
one-particle process, unless we assume very large intrachain Hubbard 
repulsion
which is out of the perturbative scheme. 
On the contrary, in the latter case, the one-particle processes are 
strongly suppressed through
 growth of the intraladder scattering processes which lead the isolated 
Hubbard ladder toward the spin gap metal phase.
Consequently when $t_{\perp}$ sets in, there exists, for any finite 
intraladder Hubbard repulsion, the region where the 
two-particle crossover dominates the one-particle crossover.

\section*{Acknowledgements}
J.K was   supported by a Grant-in-Aid for Encouragement for Young 
Scientists   from the Ministry of Education, Science, Sports and Culture, 
Japan. 

\appendix
\section{Derivation of the Scaling Equations}
The PRG program takes three steps:
1. coarse graining,  2. field rescaling and  3. 
renormalization. 

\noi\underline{\it Coarse Graining}

As depicted in Fig.~18, first    we  divide the  linearized intraladder
bands with scaling-dependent bandwidth  
$E(l)=E_{0}e^{-l}$  into
 the inner  region, $ {\cal C}_{\nu m<}$ (slow modes),  
 and the outer shell, $d{\cal C}_{\nu m>}$ (fast modes), 
where
\beq
\begin{array}{c}
{\cal C}_{\nu m<}\equiv\left\{k_{\para} \mid \mid \varepsilon_{\nu 
m}(k_{\para})\mid<E(l+dl)/2 \right\},\non\\
d{\cal C}_{\nu m>}\equiv\left\{k_{\para}  \mid E(l+dl) /2< \mid 
\varepsilon_{\nu m}(k_{\para})\mid< E(l)/2\right\}.\non
\end{array}
\eeq
Coarse graining means    integration of   the fermion  degrees of 
freedom over 
 $d{\cal C}_{\nu m>}$
with thickness 
\begin{equation}
\mid dE(l)\mid={1\over 2}E(l)dl.\non
\end{equation}
We leave $\varepsilon_{n}$ and $k_{\perp}$ free.  
This procedure is performed for each branch ($\nu m=LB,LA,RA,RB$) and spin.

The action (\ref{eqn:action}) is then decomposed into the slow and the 
fast counterparts as
 \begin{eqnarray}
S&=&S_{<}+S_{>}\non\\
&=&S^{(1)}_{\para<}+S^{(2)}_{\para<}+S^{(1)}_{\perp<}+S^{(2)}_{\perp<}+S^{(1)}_{\para>}+S^{(2)}_{\para>}+S^{(1)}_{\perp>}+S^{(2)}_{\perp>},\non
\end{eqnarray}
where $S_{<}$ consists of only the modes in $ {\cal C}_{\nu m<}$,
 while $S_{>}$ includes  the modes in $d{\cal C}_{\nu m>}$.
 The partition function is then 
rewritten as
\beq
Z=\int{\cal D}_{<}\exp\left[S^{(1)}_{\para<}+S^{(2)}_{\para<}+S^{(1)}_{\perp<}+S^{(2)}_{\perp<}\right]
\int{\cal D}_{>}\exp\left[S^{(1)}_{\para>}+S^{(2)}_{\para>}+S^{(1)}_{\perp>}+S^{(2)}_{\perp>}\right],
\label{eqn:pathinte}\non
\eeq
 where $\cal  D_{<}$ and $ \cal D_{>}$ symbolize the measure over the 
modes in 
 $ C_{\nu m<}$ and $d{\cal C}_{\nu m>}$, respectively.
Under the weak coupling condition, $t_{\perp},g^{(i)}_{\mu}\ll t,t'$,  
we expand  
$\exp\left[{S^{(1)}_{\para>}+S^{(2)}_{\para>}+S^{(1)}_{\perp>}+S^{(2)}_{\perp>}}\right]$  
with regard to 
$S^{(2)}_{\para>}$, $S^{(1)}_{\perp>}$, and $S^{(2)}_{\perp>}$,
and   integrate out the modes in $d{\cal C}_{>}$.
Then we obtain
\beq
Z=\int{\cal D}_{<}\exp\left[S^{(1)}_{\para<}+S^{(2)}_{\para<}
+S^{(1)}_{\perp<}+S^{(2)}_{\perp<}+{\sum_{l,m,n=0}^{\infty}\Gamma_{lmn}}\right]\non
,\label{eqn:avefast}
\eeq
where
\beq
\Gamma_{lmn}={{1\over l!m!n!}\langle\!\langle 
\left[S^{(2)}_{\para>}\right]^{l}\left[S^{(1)}_{\perp>}\right]^{m}\left[S^{(2)}_{\perp>}\right]^{n}
\rangle\!\rangle_{\rm c}}, \label{eqn:Cfactor}
\eeq
$
\langle\!\langle(\cdots)
\rangle\!\rangle
=\int {\cal D}_{>}e^{S^{(1)}_{\para>}}(\cdots)
$
and the subscript 'c' represents the connected diagram.
Now we obtain the  modified action, $\bar S_{<}$, 
with the modified intraladder propagators and interladder one- and 
two-particle hopping
 amplitudes 
written as
\begin{eqnarray}
\bar {\cal G}_{\nu m}^{-1}(K_{\para})&=&[1+z_{\para}(l)dl+{\cal 
O}(dl^{2})] { \cal G}_{\nu m}^{-1}(K),\label{eqn:beforefr}\\
\bar t_{\perp}&=&[1+{\cal O}(dl^{2})] t_{\perp},\\
\bar g^{(i)}_{\mu}&=&[1+z^{(i)}_{\para\mu}(l) dl+{\cal 
O}(dl^{2})]g^{(i)}_{\mu},\label{eqn:reng}\\
\bar V_{\mu}^{M}&=&f_{\mu}^{M}(l)dl+[z_{\perp\mu}^{M}(l) dl+{\cal 
O}(dl^{2})]V_{\mu}^{M},\label{eqn:renV}
\label{eqn:RGforgz}
\end{eqnarray} 
where $f_{\mu}^{M}$ symbolizes the generator of  $V_{\mu}^{M}$ which
is diagramatically shown  in Fig.~7.
In the low order arguments, $z^{M}_{\perp\mu}$ has $[V^{M}_{\mu}]^n$ contributions with $n=0$ 
and $n=1$.

We obtain the  $z_{\para}(l)$, $z^{(i)}_{\para\mu}(l)$,  $f_{\mu}^{M}(l)$
and $z_{\perp\mu}^{M}(l)$, by picking up and evaluating the Feynmann 
diagrams 
which give contributions in proportion to $dl$.
The diagrams in  Figs.4(a) and 4(b)
 give $z^{(1)}_{\para\mu}(l)$  and $z^{(2)}_{\para\mu}(l)$, respectively, 
 which come from $\Gamma_{200}+\Gamma_{300}$.
The diagrams in  Fig.4(c)  
 give $z_{\para}(l)$ which come from $\Gamma_{200}$.
The diagrams in  Fig.7  
 give $f_{\mu}^{M}$ which come from $\Gamma_{220}$.
The second and third diagrams on the r.h.s of Fig.8  
 give $z_{\perp\mu}^{M}$ which come from $\Gamma_{101}+\Gamma_{002}$.

\noi\underline{\it Field Rescaling}

The right-moving sector of the coarse-grained intraladder kinetic action  
is
\beq
\sum_{\ii 
\varepsilon_{n}}\sum_{\sigma}\int_{-E(l+dl)/2v_{F}}^{E(l+dl)/2v_{F}}{dk_{\para}\over 
2\pi }[1+z_{\para}(l)dl]
{\cal G}_{Rm}^{-1}(K_{\para})R^{\ast}_{m\sigma}(K)R_{m\sigma}(K). 
\label{eqn:bFR} 
\eeq
To restore the original cutoff and 
leave the intraladder kinetic action, $S_{\para}$, invariant, we tune the 
momenta, frequencies 
and field variables   as
\beq
\left\{\begin{array}{c}
\bar K_{\para}=e^{dl}K_{\para}=\left( 1+dl\right)K_{\para}, \\
\bar R_{m\sigma}(\bar K)=[1+{1\over 2}\{z_{\para}(l)-3\}dl]R_{m\sigma}(K),
\end{array}\right.\label{eqn:rescaling}\non
\eeq
where 
\beq
\left\{\begin{array}{c}
\bar K_{\para}=(e^{dl}k_{\para},\ii e^{dl}\varepsilon_{n}),\\
\bar K=(e^{dl} k_{\para},k_{\perp},\ii e^{dl}\varepsilon_{n}).
\end{array}\right.\non
\eeq
We must leave the transverse momentum, $k_{\perp}$, unchanged since the
scaling law is well defined  only along the longitudinal direction.
Then we  see that (\ref{eqn:bFR})  becomes
\beq
\sum_{\ii 
\bar\varepsilon_{n}}\sum_{\sigma}\int_{-E(l)/2v_{F}}^{E(l)/2v_{F}}{d\bar 
k_{\para}\over 2\pi}
{\cal G}_{Rm}^{-1}(\bar K_{\para}) 
\bar R^{\ast}_{m\sigma}(\bar K)\bar R_{m\sigma}(\bar K),\nonumber 
\eeq
which is just the same form as the initial one, namely, scale-invariant.

\noi\underline{\it Renormalization}

 Finally we renormalize the physical quantities in the action
and set up  the  differential equations for  them
with keeping $S_{\para}$ scale-invariant.  Discarding  all the 
${O}(dl^2)$ terms in (\ref{eqn:beforefr})-(\ref{eqn:RGforgz}) 
and  taking account of (\ref{eqn:RGforgz}) 
 and (\ref{eqn:rescaling}), we obtain the scaling equations
\begin{eqnarray}
{d \ln g^{(i)}_{\mu}(l) \over dl}&=&z^{(i)}_{\para\mu}(l)-2z_{\para}(l),\\
{d \ln t_{\perp}(l) \over dl}&=&1-z_{\para}(l),\\
{d  V^{M}_{\mu}(l) \over dl}&=&f_{\mu}^{M}(l)+z^{M}_{\mu}(l)V_{\mu}^{M}(l),
\end{eqnarray}
which give (\ref{eqn:generalformforg}), (\ref{eqn:RGfort}) and 
(\ref{eqn:scalingof V}), respectively.
Note that $z^{M}_{\perp\mu}$ has $[V^{M}_{\mu}]^n$ contributions with $n=0$ 
and $n=1$.

\section{Evaluation of Self-Energy Diagrams}
We here give in detail the derivation of the factor, $z_{\para}$, which 
governs the one-dimensional crossover.
In Ref.[13],  the same results are given.
The factor, $z_{\para}$, comes from the renormalization of the intraladder 
propagotor, ${\cal G}_{\nu m}(K_{\para})$.
($\nu=R,\,L:\,m=A,B$).
Since $z_{\para}$ is independent of $\nu$ and $m$, we consider only the 
case of 
${\cal G}_{RB}(K_{\para})$.

The 2nd order renomalization   of ${\cal G}_{\nu m}(K_{\para})$ originates 
from
$
\Gamma_{200}={1\over 2}\langle\langle \left[S^{(2)}_{\para>}\right]^2 
\rangle\rangle.
$
In Fig.19, we give diagrammatic representations for $\Gamma_{200}$ 
which renoremalize ${\cal G}_{RB}(K_{\para})$.
Although  there are a lot of ways to assign the outer-shell mode in  
$S^{(2)}_{\para}$,
there are only two possible ways which  renormalize
 ${\cal G}_{\nu m}(K_{\para})$.
We specify these two cases by arrows in Fig.19:
\[
\left\{\begin{array}{c}
{\rm case\!-\!I}:\,\,\,k'_{\para}\in d{\cal 
C}_{Lm_{1}>},\,\,k'_{\para}+q_{\para}\in {\cal 
C}_{Lm_{2}<},\,\,k_{\para}-q_{\para}\in {\cal C}_{Rm_{3}<},\non\\
{\rm case\!-\!I\!I}:\,\,\,k'_{\para}\in {\cal C}_{Rm_{1}<} 
,\,\,\,k'_{\para}+q_{\para}\in d{\cal 
C}_{Lm_{2}>},\,\,k_{\para}-q_{\para}\in {\cal C}_{Rm_{3}<}.\non
\end{array}\right.
\]
Then, among a lot of terms included in  $\Gamma_{200}$, the contribution to the 
renormalization of ${\cal G}_{RB}(k_{\para})$ 
 is  obtained  as
\begin{eqnarray*}
\Gamma^{\rm S.E}_{200}&=&
2\left(2\pi v_{F} \right)^2\left[
     ({{ g}_{0}^{(1)}}^2
    +{{ g}_{0}^{(2)}}^2-{ g}_{0}^{(1)}
     { g}_{0}^{(2)}){\cal I}^{BBB}(dl)
\right.\non\\
 && +{{ g}_{f}^{(1)}}^2{\cal I}^{BAA}(dl)
    +{{ g}_{f}^{(2)}}^2{\cal I}^{AAB}(dl)
-{ g}_{f}^{(1)}{ g}_{f}^{(2)}{\cal I}^{AAB}(dl)\non\\
  &&  
+\left.{{ g}_{t}^{(1)}}^2{\cal I}^{AAB}(dl)
+{{ g}_{t}^{(2)}}^2{\cal I}^{BAA}(dl)
-{ g}_{t}^{(1)}{ g}_{t}^{(2)}{\cal I}^{BAA}(dl)\right],
\end{eqnarray*}
where 
\[{\cal I}^{m_{1}m_{2}m_{3}}(dl)={\cal I}^{m_{1}m_{2}m_{3}}_{\rm 
I}(dl)+{\cal I}^{m_{1}m_{2}m_{3}}_{\rm I\!I}(dl)\] 
with
${\cal I}^{m_{1}m_{2}m_{3}}_{\rm I}(dl)$ and ${\cal 
I}^{m_{1}m_{2}m_{3}}_{\rm I\!I}(dl)$ corresponding to the cases I and $\rm 
I\!I$, respectively.
${\cal I}^{m_{1}m_{2}m_{3}}_{\rm I}(dl)$ is evaluated as
\begin{eqnarray*}
{\cal I}^{m_{1}m_{2}m_{3}}_{\rm I}(dl)&=&
T^{2}\ds\int_{dC_{Lm_{1}>}} {d k'_{\para}\over 2\pi}\int {dq\over 
2\pi}\sum_{m,l}
{\cal G} _{Lm_{1}}(k',\ii\varepsilon'_{m})
{\cal G} _{Rm_{2}}(k-q,\ii \varepsilon_{n}-\ii\omega_{l})
{\cal G} _{Lm_{3}}(k'+q,\ii\varepsilon'_{m}+\ii\omega_{l})\\
&=&-{1\over 4}\ds\int_{dC_{Lm_{1}>}} {d \varepsilon'_{Lm_{1}}\over 2\pi 
v_{F}}\int {dq\over 2\pi}
{{\cal T}(\varepsilon'_{Lm_{1}},k,q;T)
\over D+2v_{F}(q-\Delta k_{F})},
\end{eqnarray*}
where $\Delta k_{F}=k_{Fm_{1}}-k_{Fm_{2}}$ and $D=\ii 
\varepsilon_{n}-v_{F}(k-k_{FB})$.
The thermal factor is given by 
\begin{eqnarray*}
&&{\cal T}(\varepsilon_{Lm_{1}},K,q;T)=\\
&&\left[
\tanh{\varepsilon_{Lm_{1}}\over 2T}-
\tanh{\varepsilon_{Lm_{1}}+v_{F}(\Delta k_{F}-q)\over 2T}\right]
\left[\coth{v_{F}(q-\Delta k_{F})\over 2T}-\tanh{v_{F}(k-k_{FB}-q+\Delta 
k_{F})\over 2T}\right].
\end{eqnarray*}
In the low-temperature limit, 
\[{\cal T}(\varepsilon_{Lm_{1}},k,q;T=0)=\left\{\begin{array}{c}
4\,\,\,\,\,\,{\rm 
for}\,\,\,\,\,\varepsilon_{Lm_{1}}\left[\varepsilon_{Lm_{1}}+v_{F}(\Delta 
k_{F}-q)\right]<0,\\
0\,\,\,\,\,\,{\rm 
for}\,\,\,\,\,\varepsilon_{Lm_{1}}\left[\varepsilon_{Lm_{1}}+v_{F}(\Delta 
k_{F}-q)\right]>0.
\end{array}\right.\]
Taking account of  the restriction on available $q_{\para}$, coming from 
the case-I condition,   we obtain
\begin{eqnarray*}
{\cal I}^{m_{1}m_{2}m_{3}}_{\rm I}(dl)
&=&-{1\over 4\pi^2 v_{F}}\left[\ds\int_{E_{0}-\mid d E_{0}(l) \mid\over 
2}^{E_{0}\over 2}d\varepsilon'_{Lm_{1}}
\int_{{E_{0}\over 2v_{F}}+\Delta k_{F}}^{{E_{0}\over v_{F}}+\Delta k_{F}}dq
+\ds\int_{-E_{0}\over 2}^{-{E_{0}-\mid d E_{0}(l) \mid\over 
2}}d\varepsilon'_{Lm_{1}}
\int_{-{E_{0}\over v_{F}}+\Delta k_{F}}^{-{E_{0}\over 2v_{F}}+\Delta 
k_{F}}dq\right]\non\\
&&{1\over D+2v_{F}(q-\Delta k_{F})} \nonumber\\
&=&
=-{ \mid d E_{0}(l) \mid\over 16\pi^2 v_{F}^{2}}
\ds
\ln{
\left[  E_{0}-D\right]
\left[2E_{0}+D\right]
\over 
\left[E_{0}+D\right]
\left[2E_{0}- D\right]}.
\end{eqnarray*}
By taking the limit  
$ D/ E_{0}\ll 1$,
we obtain
\[
{\cal I}^{m_{1}m_{2}m_{3}}_{\rm I}(dl)=
{D\over 16\pi^2 v_{F}^{2}}dl
={1\over 16\pi^2 v_{F}^{2}} {\cal G}_{RB}^{-1}(K)dl.
\]
Evaluation of ${\cal I}^{m_{1}m_{2}m_{3}}_{\rm I\!I}(dl)$ gives the same 
result.
Thus we obtain
\[
\Gamma^{\rm S.E}_{200}=z_{\para}{\cal G}_{RB}^{-1}(K)dl,
\]
and the modified intraladder one-particle action  
\begin{eqnarray*}
\bar S^{(1)}_{\para}=\sum_{K} 
\sum_{m=A,B}\sum_{\sigma}(1+z_{\para}dl)\left[{\cal G}_{Lm}^{-1}(K_{\para})
L^{\ast}_{m\sigma}(K)L_{m\sigma}(K)\right.
+\left.{\cal G}_{Rm}^{-1}(K_{\para})
R^{\ast}_{m\sigma}(K)R_{m\sigma}(K)
\right],
\end{eqnarray*}
where $z_{\para}$ is given by (\ref{eqn:zpara}).
This result leads to (\ref{eqn:beforefr}).

\section{Evaluation of Generators for Interladder Two-Particle Hopping 
Processes}
We here give a full account of evaluation of the generators for the
two-particle hopping processes, $f_{\mu}^{M}$.
It is convenient to rewrite the intraladder 
interaction part
in terms of the composite particle variables as
\begin{eqnarray}
S_{\rm int}&=&
{\pi v_{F}\over 2}\sum_{Q_{\para}}\left[
g^{\rm CDW}_{0}{\cal O}^{BB\ast}_{\rm CDW}{\cal O}^{BB}_{\rm CDW}\right.
+g^{\rm CDW}_{t}{\cal O}^{BA\ast}_{\rm CDW}{\cal O}^{AB}_{\rm CDW}
+g^{\rm CDW}_{f}{\cal O}^{BA\ast}_{\rm CDW}{\cal O}^{BA}_{\rm CDW} \non\\
&+&g^{\rm SDW}_{0} \vec{\cal O}^{BB\ast}_{\rm SDW}\cdot\vec{\cal O}^{BB}_{\rm 
SDW} 
+g^{\rm SDW}_{t} \vec{\cal O}^{BA\ast}_{\rm SDW}\cdot\vec{\cal O}^{AB}_{\rm 
SDW}
+g^{\rm SDW}_{f} \vec{\cal O}^{BA\ast}_{\rm SDW}\cdot\vec{\cal O}^{BA}_{\rm 
SDW}\\
&+&g^{\rm SS}_{0}{\cal O}^{BB\ast}_{\rm SS}{\cal O}^{BB}_{\rm SS}  
+g^{\rm SS}_{t}{\cal O}^{BB\ast}_{\rm SS}{\cal O}^{AA}_{\rm SS}
+g^{\rm SS}_{f}{\cal O}^{AB\ast}_{\rm SS}{\cal O}^{AB}_{\rm SS} \non\\
&+&g^{\rm TS}_{0}\vec{\cal O}^{BB\ast}_{\rm TS}\cdot\vec{\cal O}^{BB}_{\rm 
TS}
+g^{\rm TS}_{t}\vec{\cal O}^{BB\ast}_{\rm TS}\cdot\vec{\cal O}^{AA}_{\rm 
TS} 
+\left.g^{\rm TS}_{f}\vec{\cal O}^{AB\ast}_{\rm TS}\cdot\vec{ \cal
O}^{AB}_{\rm TS}
\right]+(A\leftrightarrow B),\non \label{eqn:acom}
\end{eqnarray}
where 
${\cal O}^{m_{1}m_{2} }_{M}$ denotes ${\cal O }^{m_{1}m_{2} 
}_{M}(Q_{\para})$.

As an illustration,  we derive the   generator for the CDW 
channel, $f_{\mu}^{\rm CDW}$. 
In the action, (\ref{eqn:acom}), the CDW channel consists of
\begin{eqnarray*}
S_{\rm int }^{\rm CDW}&=&
{1\over 2}\pi v_{F}g^{\rm CDW}_{0}\sum_{Q_{\para}} 
{\cal O}^{BB\ast}_{\rm CDW}(Q_{\para}){\cal O}^{BB}_{\rm CDW}(Q_{\para})\\
&+&
{1\over 2}\pi v_{F}\sum_{Q_{\para}}\left[
 g^{\rm CDW}_{t}{\cal O}^{BA\ast}_{\rm CDW}(Q_{\para}){\cal O}^{AB}_{\rm 
CDW}(Q_{\para})
+g^{\rm CDW}_{f}{\cal O}^{BA\ast}_{\rm CDW}(Q_{\para}){\cal O}^{BA}_{\rm 
CDW}(Q_{\para})\right]
+(A\leftrightarrow B).
\end{eqnarray*}
The   generators depicted in Fig.~7(a)  come from 
$\Gamma_{022}$.
Among a lot of terms included in  $\Gamma_{220}$, the contribution 
to $f_{\mu}^{\rm CDW}$, $\Gamma_{220}^{{\rm CDW}}$, is
 obtained by  assigning   composite field  variables to the outer
  shell mode as
\begin{eqnarray*}
S_{\rm int > }^{\rm CDW}&=&
{1\over 2}\pi v_{F}g^{\rm CDW}_{0}\textstyle\sum_{Q_{\para}}
({\cal O}^{BB\ast}_{\rm CDW>}{\cal O}^{BB}_{\rm CDW<}
+{\cal O}^{BB\ast}_{\rm CDW<}{\cal O}^{BB}_{\rm CDW>}
+{\cal O}^{AA\ast}_{\rm CDW>}{\cal O}^{AA}_{\rm CDW<}
+{\cal O}^{AA\ast}_{\rm CDW<}{\cal O}^{AA}_{\rm CDW>})\\
&+&
{1\over 2}\pi v_{F}\textstyle\sum_{Q_{\para}}[
 g^{\rm CDW}_{t}({\cal O}^{BA\ast}_{\rm CDW>}{\cal O}^{AB}_{\rm CDW<}
+{\cal O}^{BA\ast}_{\rm CDW<}{\cal O}^{AB}_{\rm CDW>}  
 +{\cal O}^{AB\ast}_{\rm CDW>}{\cal O}^{BA}_{\rm CDW<}
+{\cal O}^{AB\ast}_{\rm CDW<}{\cal O}^{BA}_{\rm CDW>})\\
&+&g^{\rm CDW}_{f} ({\cal O}^{BA\ast}_{\rm CDW>}{\cal O}^{BA}_{\rm CDW<}
+{\cal O}^{BA\ast}_{\rm CDW<}{\cal O}^{BA}_{\rm CDW>}+{\cal O}^{AB\ast}_{\rm CDW>}
{\cal O}^{AB}_{\rm CDW<}+{\cal O}^{AB\ast}_{\rm CDW<}{\cal O}^{AB}_{\rm 
CDW>})
]\\
&&+(A\leftrightarrow B),
\end{eqnarray*}
where
\begin{eqnarray*}
{\cal O}^{mm'}_{\rm CDW<}&=&
\beta^{-1/2}
\sum_{k_{\para}+q_{\para}\in {\cal C}_{Rm<}}
\sum_{k_{\para}\in {\cal C}_{Lm<}}
\sum_{\ii\varepsilon_{n}}\sum_{\sigma}
R^{*}_{m\sigma}(K_{\para}+Q_{\para})L_{m'\sigma}(K_{\para}),\\
{\cal O}^{mm'}_{\rm CDW>}&=&
\beta^{-1/2}
\sum_{k_{\para}+q_{\para}\in d{\cal C}_{Rm>}}
\sum_{k_{\para}\in d{\cal C}_{Lm>}}
\sum_{\ii\varepsilon_{n}}\sum_{\sigma}
R^{*}_{m\sigma}(K_{\para}+Q_{\para})L_{m'\sigma}(K_{\para}).
\non\end{eqnarray*}
Then we obtain
\begin{eqnarray*}
&&\Gamma_{220}^{\rm CDW}={1\over 4} \langle\langle S_{\perp>}^{2} 
\{S^{\rm CDW}_{\rm int>}\}^{2}\rangle\rangle\\
&=&{\pi^2 v_{F}^2\over 8}(g^{\rm CDW}_{0})^{2}\sum_{Q} 
\langle\langle S_{\perp>}^{2}{\cal O}^{BB\ast}_{\rm CDW>}{ O}^{BB}_{\rm 
CDW>} \rangle\rangle
{\cal O}^{BB\ast}_{\rm CDW<}{\cal O}^{BB}_{\rm CDW<}\\
&+&{\pi^2 v_{F}^2\over 8}\sum_{Q}\left[(g^{\rm CDW}_{t})^{2}\langle\langle 
S_{\perp>}^{2} {\cal O}^{AB\ast}_{\rm CDW>} {\cal O}^{AB}_{\rm CDW>} 
\rangle\rangle
+(g^{\rm CDW}_{f})^{2}\langle\langle S_{\perp>}^{2}  {\cal O}^{BA\ast}_{\rm 
CDW>}{\cal O}^{BA}_{\rm CDW>} \rangle\rangle
\right]{\cal O}^{BA\ast}_{\rm CDW<}{\cal O}^{BA}_{\rm CDW<}\\
&+&{\pi^2 v_{F}^2\over 4}g^{\rm CDW}_{t}g^{\rm 
CDW}_{f}\sum_{Q}\langle\langle S_{\perp>}^{2} {\cal O}^{AB\ast}_{\rm CDW>} 
{\cal O}^{AB}_{\rm CDW>} \rangle\rangle
{\cal O}^{BA\ast}_{\rm CDW<}{\cal O}^{AB}_{\rm CDW<}+(A\leftrightarrow B),
\end{eqnarray*}
where the first, second and third lines on the r.h.s of the second equation
correspond to the first, second and third lines of Fig.~7(a).
We obtain, in the low temperature limit,  
\begin{eqnarray*}
&&\langle\langle  {\cal O}^{mm}_{\rm CDW>}S_{\perp}^{2}{\cal O}^{mm\ast}_{\rm 
CDW>}\rangle\rangle
=-t_{\perp}^{2}\cos( q_{\perp})
T\sum_{k_{\para}\in d{\cal C}_{Rm>}}
\sum_{\varepsilon_{n}}
{ G} _{Lm}^{2}(k_{\para}-2k_{Fm},\ii\varepsilon_{n})
{ G} _{Rm}^{2}(k_{\para},\ii\varepsilon_{n})\non\\
&=&{1\over 2\pi v_{F}}t_{\perp}^{2}\cos( q_{\perp}d_{\perp})
\left[\ds\int_{-{E_{0}(l)\over 2}}^{-{E_{0}(l)-\mid d E_{0}(l) \mid\over 
2}}+
\ds\int_{{E_{0}(l)-\mid d E_{0}(l) \mid\over 2}}^{{E_{0}(l)\over 
2}}\right]d\varepsilon_{Rm}
\left[{{\beta\over 2}\coth^{-2}{\beta\varepsilon_{RB}\over 2}\over 
(-2\varepsilon_{RB} )^{2}}
+{2\tanh{\beta\varepsilon_{R}\over 2}\over (-2\varepsilon_{RB})^3}\right]\\
&=&-{t_{\perp}^{2}dl\over \pi v_{F}}E_{0}^{-2}(l)\cos q_{\perp}.
\end{eqnarray*}

Similar manipulation gives  
\[
\langle\langle   {\cal O}^{m\bar m}_{\rm DW}S_{\perp}^{2}{\cal O}^{m\bar 
m\ast}_{\rm 
DW}\rangle\rangle
=+{t_{\perp}^{2}dl\over \pi v_{F}}E_{0}^{-2}(l)\cos q_{\perp},
\]
where $m$ and $\bar m$ denote different bands.
Thus   we obtain
\begin{eqnarray*}
\Gamma_{220}^{\rm CDW}
&=&-{t_{\perp}^{2}\pi v_{F}\over 8E_{0}^{2}(l)}dl(g^{\rm 
CDW}_{0})^{2}\sum_{Q} 
\cos q_{\perp}
{\cal O}^{BB\ast}_{\rm CDW<}{\cal O}^{BB}_{\rm CDW<}\\
&+&{t_{\perp}^{2}\pi v_{F}\over 8E_{0}^{2}(l)}dl\left[(g^{\rm CDW}_{t})^{2}
+(g^{\rm CDW}_{f})^{2}\right]
\sum_{Q_{\para}}\cos q_{\perp}{\cal O}^{BA\ast}_{\rm CDW<}{\cal O}^{BA}_{\rm CDW<}\\
&+&{t_{\perp}^{2}\pi v_{F}\over 4E_{0}^{2}(l)}dlg^{\rm CDW}_{t}g^{\rm 
CDW}_{f}
\sum_{Q_{\para}}\cos q_{\perp}
{\cal O}^{BA\ast}_{\rm CDW<}{\cal O}^{AB}_{\rm CDW<}.
\end{eqnarray*}
After  the field rescaling procedure, the first, second and third lines of the above equation give the first terms in 
(\ref{eqn:CDW1}), (\ref{eqn:CDW2}) and (\ref{eqn:CDW3}), respectively.
It is here useful to present the following results.
\begin{eqnarray*}
\begin{array}{c}
\langle\langle  {\cal O}^{mm}_{\rm DW}S_{\perp}^{2}{\cal O}^{mm\ast}_{\rm 
DW}\rangle\rangle
=-{t_{\perp}^{2}dl\over \pi v_{F}}E_{0}^{-2}(l)\cos q_{\perp},\\
\langle\langle   {\cal O}^{m\bar m}_{\rm DW}S_{\perp}^{2}{\cal O}^{m\bar 
m\ast}_{\rm DW}\rangle\rangle
=+{t_{\perp}^{2}dl\over \pi v_{F}}E_{0}^{-2}(l)\cos q_{\perp},\\
\langle\langle  {\cal O}^{mm}_{\rm S}S_{\perp}^{2}{\cal O}^{mm\ast}_{\rm 
S}\rangle\rangle
=+{t_{\perp}^{2}dl\over \pi v_{F}}E_{0}^{-2}(l)\cos q_{\perp},\\
\langle\langle   {\cal O}^{m\bar m}_{\rm S}S_{\perp}^{2}{\cal O}^{m\bar 
m\ast}_{\rm S}\rangle\rangle
=-{t_{\perp}^{2}dl\over \pi v_{F}}E_{0}^{-2}(l)\cos q_{\perp},
\end{array}
\end{eqnarray*}
where DW=CDW/SDW and S=SS/TS.
When we evaluate the second and third diagrams of Figs.8(a) and 8(b), it is     
useful to note the
 outer shell integration, which appears  in the evaluation of $ \Gamma_{011}$ 
 and $ 
\Gamma_{002}$, is given by
\begin{eqnarray*}
\langle\langle  {\cal O}^{mm}_{M}{\cal O}^{mm\ast}_{M}\rangle\rangle
=\langle\langle   {\cal O}^{m\bar m}_{M}{\cal O}^{m\bar 
m\ast}_{M}\rangle\rangle
={dl\over \pi v_{F}}
\end{eqnarray*}
for $M=$ CDW, SDW, SS or TS.


\pagebreak
\noi
Fig.~1: Array of Hubbard ladders studied here. 

\bigskip\noi Fig.~2: Schematic illustrations of the one-particle process and the  
two-particle process (in   the $d$-wave superconductivity channel).
In the one-particle process, a particle  hops  from one ladder to a 
neighboring one, while
in the two-particle process,
 a pair of particles (bipolaron) hops  from one ladder to a 
neighboring one.

\bigskip\noi Fig.~3: (a) Four branches($LB,LA,RA,RB$) of  linearized bands with the 
bandwidth cutoff $E$, and 
(b) intraladder two-particle scattering vertices $g^{(i)}_{\mu}$. The 
solid and broken lines represent the propagators for the right-moving and 
left-moving electrons, respectively. $m$ and $\bar m$ denote different bands.

\bigskip\noi Fig.~4: Diagrams which contribute to   
 the 3rd order vertex corrections,
(a) to   the intraladder backward scattering processes $g_{\mu}^{(1)}$,   
(b) to   the intraladder forward  scattering processes $g_{\mu}^{(2)}$,  
and   
(c) to the intraladder self-energy processes.
Wavy lines represent one of the intraladder scattering vertices, 
$g_{\mu}^{(i)}$ ($i=1,\,2;\,\,\mu=0,\,f,\,t$).

\bigskip\noi Fig.~5: Diagrammatic representation of the scaling equation for the 
interladder one-particle hopping amplitude.
A zigzag line represents the interladder one-particle hopping, 
$t_{\perp}$.

\bigskip\noi Fig.~6: Scaling flows of the interladder one-particle hopping amplitude, 
$t_{\perp}$, 
(a) for $\tilde U=0.3$, and (b) for $\tilde t_{\perp 0}=0.01$.
For smaller $\tilde U$ and larger $\tilde t_{\perp 0}$, $\tilde 
t_{\perp}(l)$ exceeds unity in the course of the scaling.

\bigskip\noi Fig.~7: Diagrammatic representations of the generators of the interladder 
two-particle hopping processes, $f_{\mu}^{M}$, 
(a) for the density wave channels,
and 
(b) for the superconducting channels. The first, second and third lines 
correspond to the flavor indices $\mu=0,f$ and $t$, respectively.
 Hatched circles  represent the 
coupling strengths, $g_{\mu}^{M}$.

\bigskip\noi Fig.~8: Diagrammatic representations of the leading order scaling 
equations for the two-particle hopping amplitudes, $V_{\mu}^{M}$, (a) for the 
density wave channels ($M=$CDW or SDW),
and 
(b) for the superconducting channels ($M=$SS or TS). 
Hatched squares represent $V_{\mu}^{M}$.

\bigskip\noi Fig.~9: (a), (b),  (c) Scaling flows of 
the intraladder coupling strengths for the composite fields, 
$g_{\mu}^{M}$, 
and 
(d), (e), (f) scaling flows of the interladder two-particle hopping amplitudes, $V^{M}_{\mu}$,  for 
$\tilde U=0.2,\,0.3,\,0.4$, respectively,  and $\tilde t_{\perp 0}=0.01$.
The vertical broken line corresponds to the scaling parameter, $l_{c}$, at 
which $V^{{\rm SC}d}$ diverges.

\bigskip\noi Fig.~10: Scaling flows of  the stiffness of the total-spin mode, 
$K_{\sigma+}$,  the interladder one-particle hopping amplitude, $\tilde 
t_{\perp}$, and  
the interladder two-particle hopping amplitude in the SCd channel, 
$V^{{\rm SCd}}$, for   various initial conditions ($\tilde U, \tilde 
t_{\perp0}$).
 For the region, $l>{\rm  Min}(l_{\rm c},l_{\rm cross})$, the scaling 
 flows, drawn by  broken curves, have no physical meaning, since the 
weak coupling picture breaks down in the 
region.

\bigskip\noi Fig.~11: Phase diagram of the weakly coupled Hubbard ladder system 
spanned by $\tilde t_{\perp0}$ and the reduced temperature $\tilde 
T=T/E_{0}$
for  $\tilde U=0.3$.
{\bf SGM}, {\bf SCd} and {\bf 2D} denote the sping gap metal phase,
the $d$-wave superconducting phase and the two-dimensional phase, respectively.
Gradual chanage of   darkness in the SGM phase schematically 
depicts the gradual approach of the isolated systems to their low-energy 
asymptotics.
Thick broken lines denote the crossover boundaries.

\bigskip\noi Fig.~12: Phase diagram of the weakly coupled Hubbard ladder system 
spanned by the intraladder Hubbard repulsion $\tilde U$ and the reduced temperature 
$\tilde T=T/E_{0}$
for  $\tilde t_{\perp0}=0.01$.
Gradual chanage of  darkness in the SGM phase schematically 
depicts the gradual approach of the isolated systems to their low-energy 
asymptotics.Thick broken lines denote the crossover boundaries.

\bigskip\noi Fig.~13: Dependence of the crossover value of the interladder one-particle 
hopping amplitude, $\tilde t_{\perp c}$, on the
intraladder Hubbard repulsion, $\tilde U$.
For $\tilde t_{\perp0}>\tilde t_{\perp c}$, the system undergoes the crossover to the 
2D phase, while
for  $\tilde t_{\perp0}<\tilde t_{\perp c}$, the SGM phase transits to the 
SCd phase.

\bigskip\noi Fig.~14: (a) Coupled ladder system with
the misfit of the rungs in the neighboring ladders  which actually exists  
in real Sr$_{14-x}$Ca$_{x}$Cu$_{24}$O$_{41}$ compounds
 and
(b) static interband polarization, $\chi_{AB}(q_{\para},q_{\perp})$, on the 1st Brillouin zone along the line $\Gamma(0,0)\to {\rm X}(\pi,0) 
\to {\rm M} (\pi,\pi) \to \Gamma$
for $t''/t=0,0.2,0.3$ and temperature, $T=10^{-5}t$. 

\bigskip\noi Fig.~15: (a) Two branches($L,R$) of  linearized bands with the bandwidth 
cutoff $E$, and 
(b) intrachain two-particle scattering vertices $g^{(i)}$. The 
solid and broken lines represent the propagators for the right-moving and 
left-moving electrons, respectively.

\bigskip\noi Fig.~16: Scaling flows of  the interchain one-particle hopping amplitude, $\tilde t_{\rm chain 
\perp}$  (the upper half plane ),   and 
the interchain two-particle hopping amplitude in the SDW channel,  $ 
V^{\rm SDW}$  (the lower half plane ),  for 
$\tilde U=0.1,\,0.2,\,0.3$,  and $\tilde t_{\perp 0}=0.01$.

\bigskip\noi Fig.~17: Phase diagrams of the weakly coupled chains,
(a) for
$\tilde U=0.3$, and 
(b) for $\tilde t_{\rm chain \perp0}=0.01$.
{\bf TL}  and {\bf 2D} denote the Tomonaga-Luttinger liquid phase and
 the two-dimensional phase, respectively.
Thick broken lines denote the crossover boundaries.

\bigskip\noi Fig.~18: Division of the four branches of the linearized bands into the 
inner region, $ C_{\nu m<}$,  and the outer shell, $d{\cal C}_{\nu m>}$.

\bigskip\noi Fig.~19: All labeled diagrams which renormalize ${\cal G}_{RB}(K_{\para})$.
Possible combinations of the band indices are 
$m_{1}m_{2}m_{3}=BBB,ABA,BBA$.
Inserted arrows with the labels I and I$\!$I denote the outer-shell modess 
in the cases I and I$\!$I, respectively.


\begin{thebibliography}{999}
\bibitem{comment1} We here ignore the misfit of the rungs in the 
neighboring ladders  which  exists  in real 
Sr$_{14-x}$Ca$_{x}$Cu$_{24}$O$_{41}$
compounds, since   we  take the continuum limit along the leg and 
consequently the misfit  never appears in the present theory.
\bibitem{comment2}In the case of the 2nd order RG, the strong coupling 
values, $+1$ and $-1$, are replaced with $+\infty$ and $-\infty$, 
respectively.
\end{thebibliography}
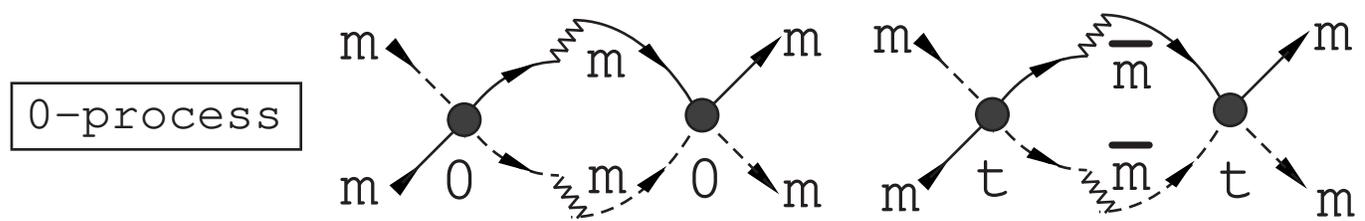
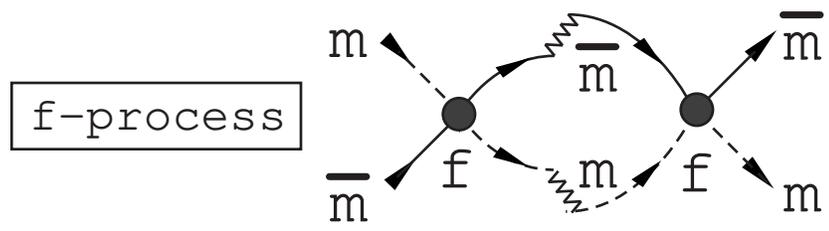
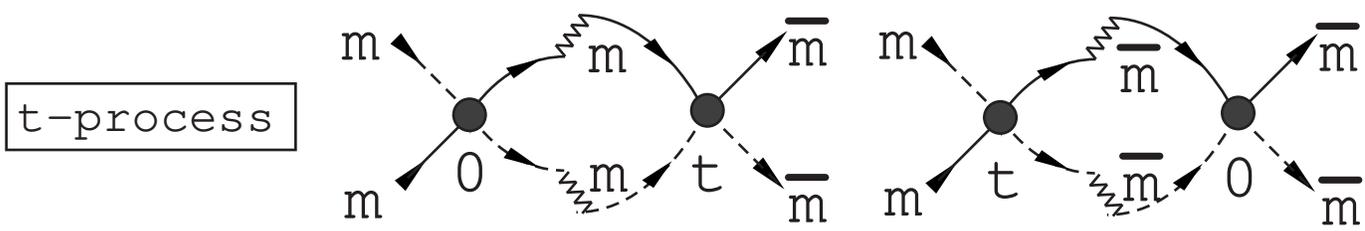
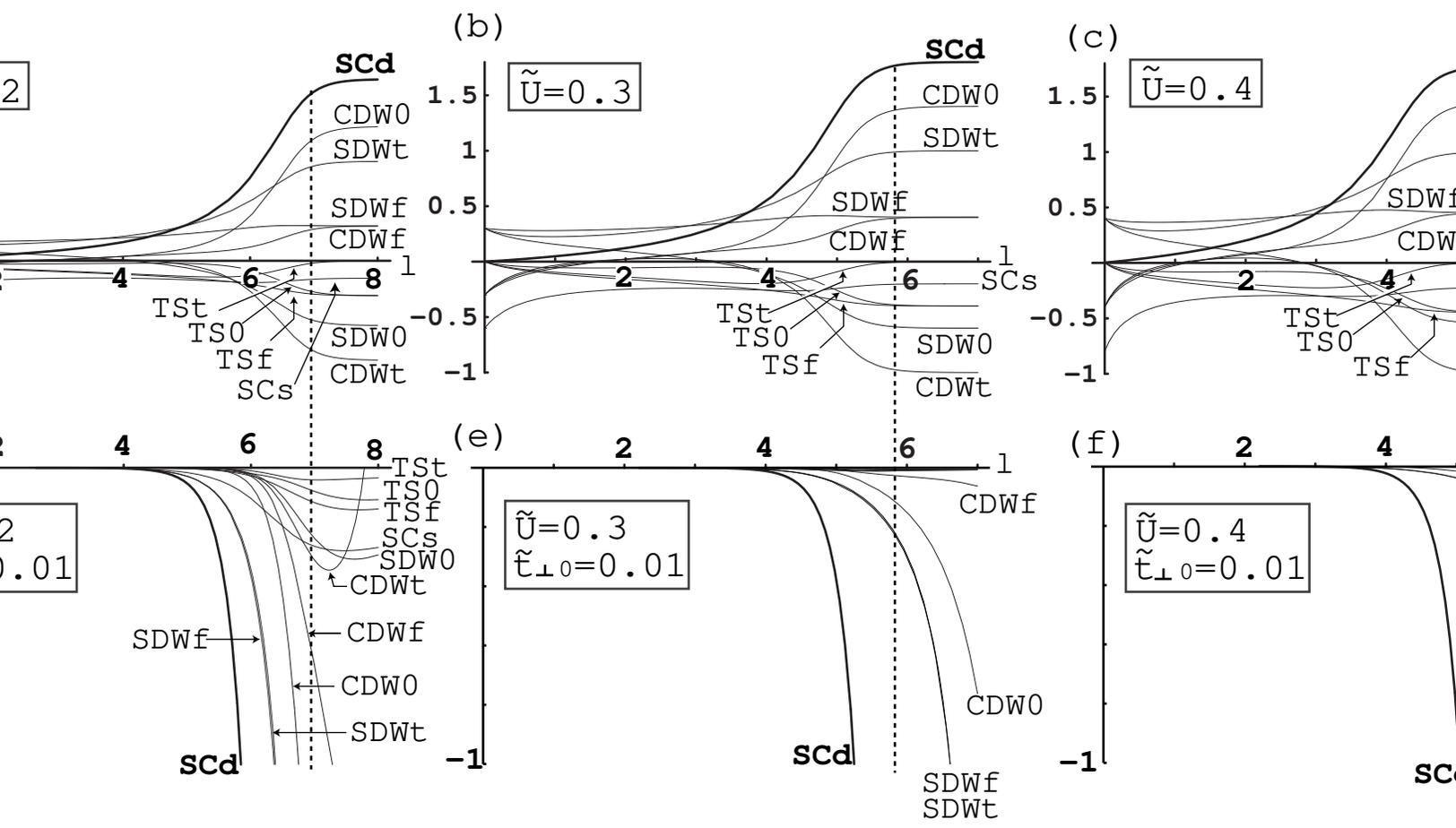
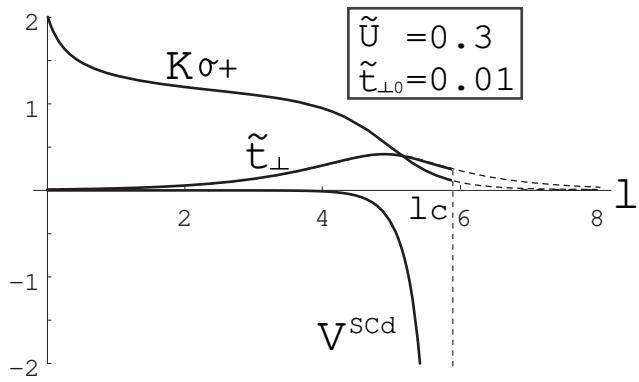
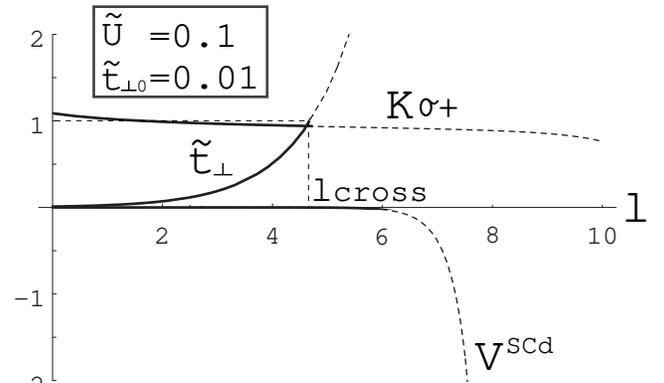
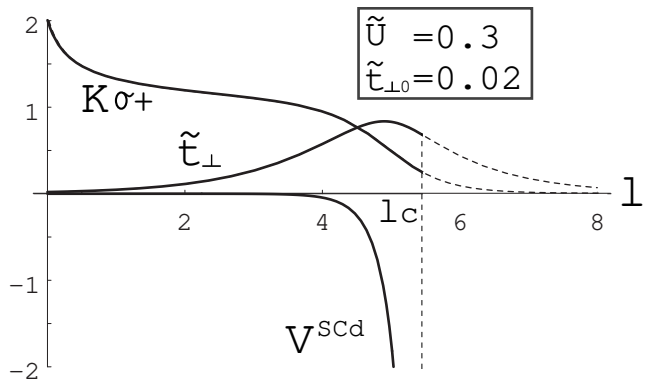
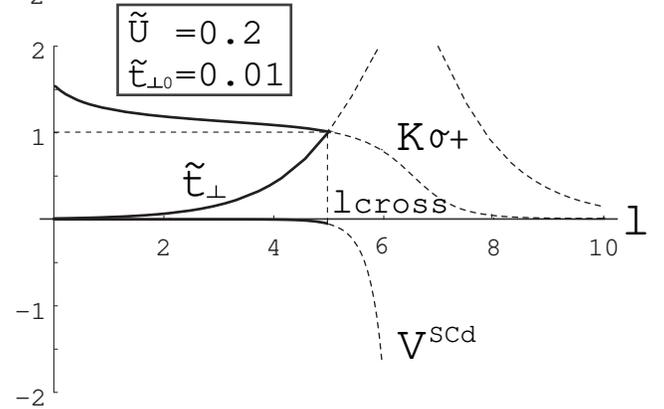
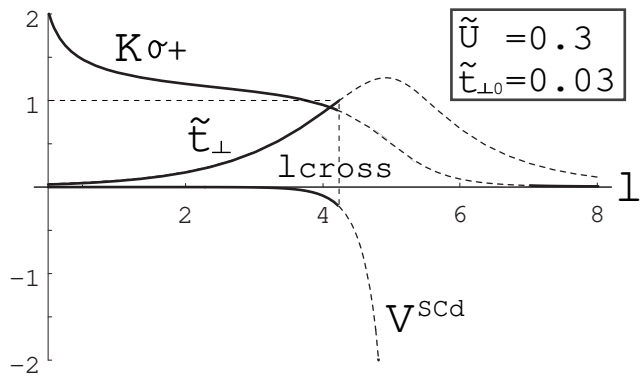
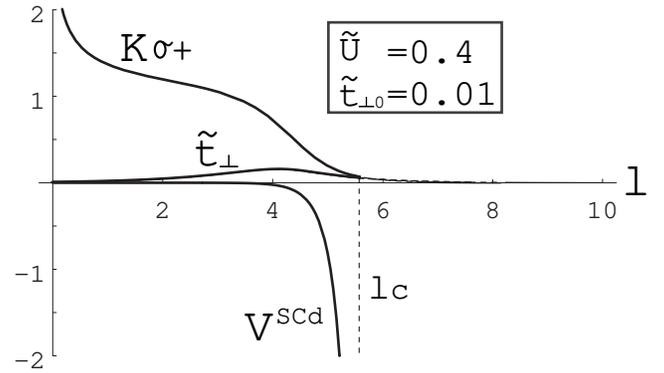
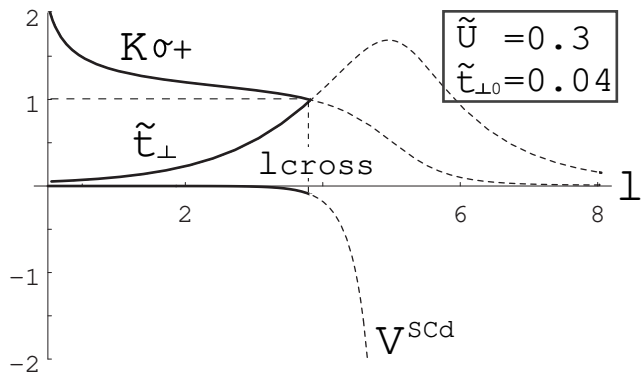
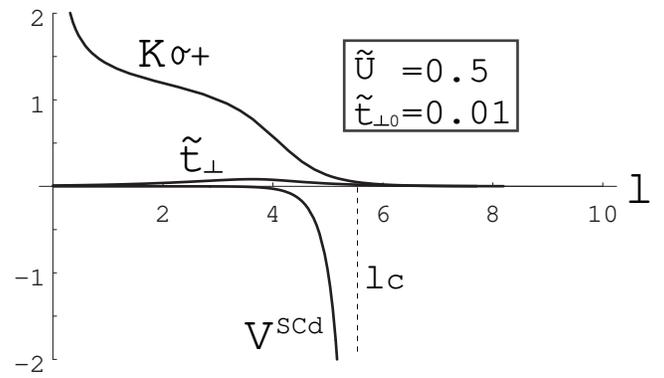
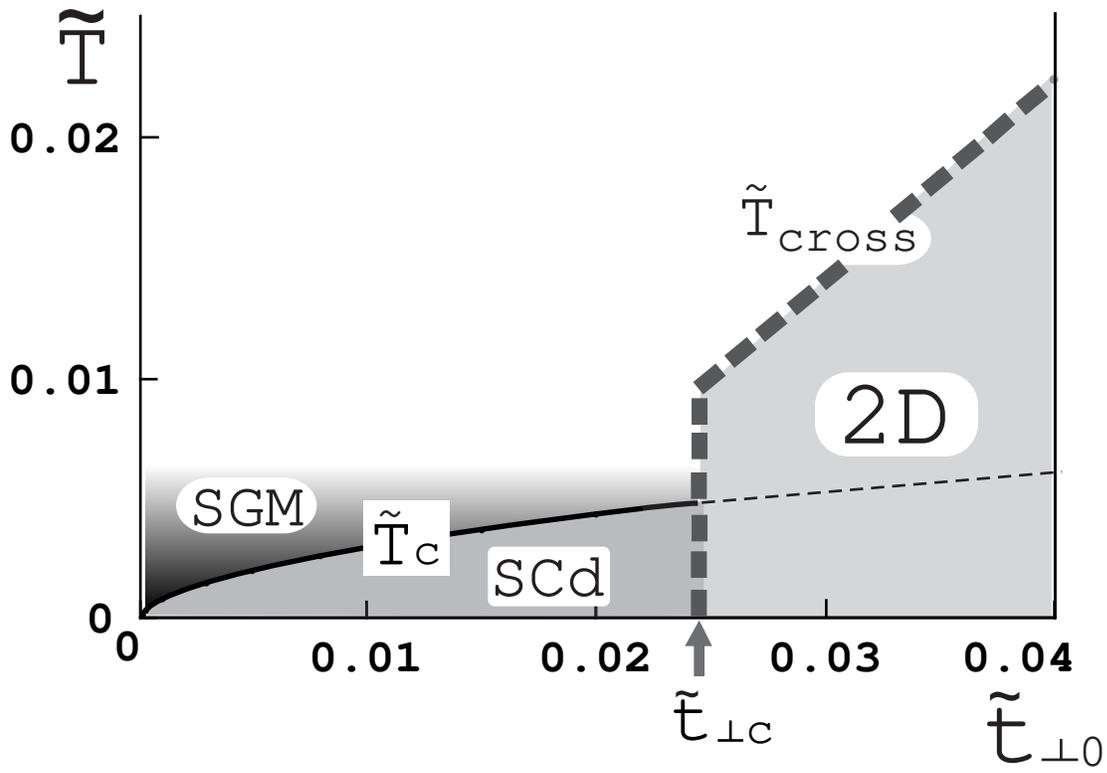
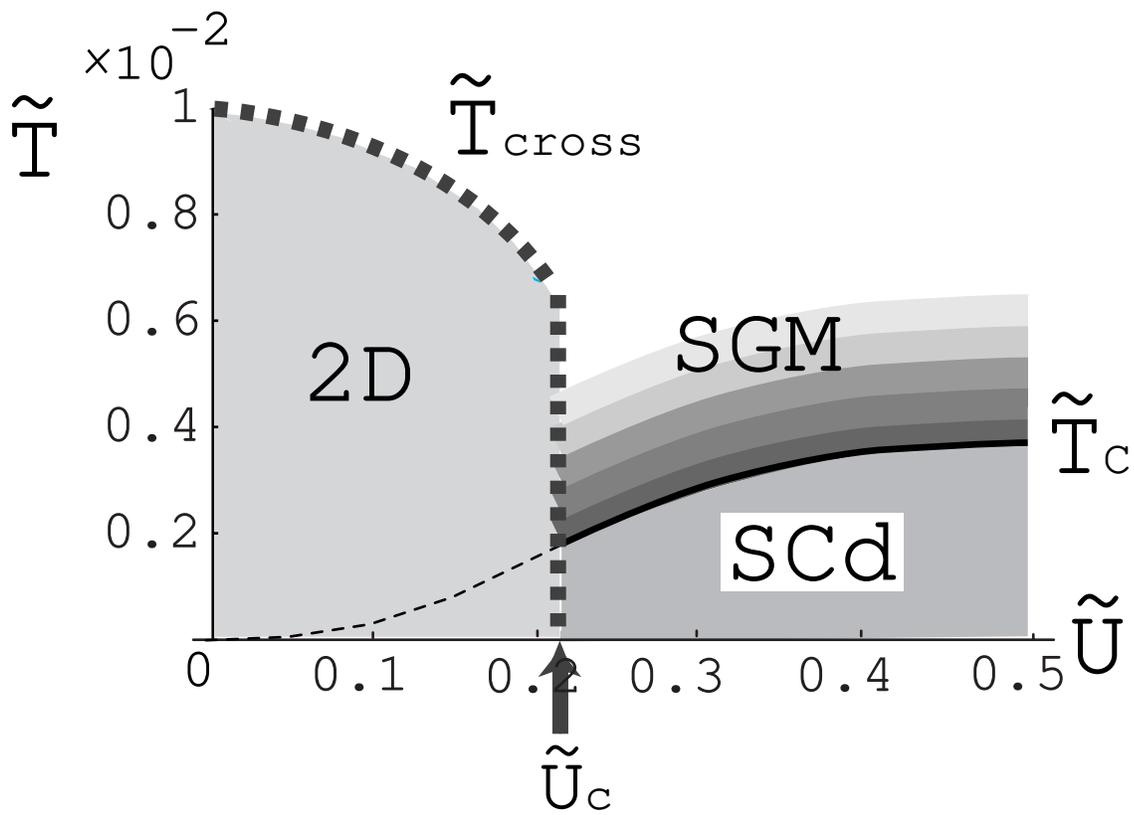
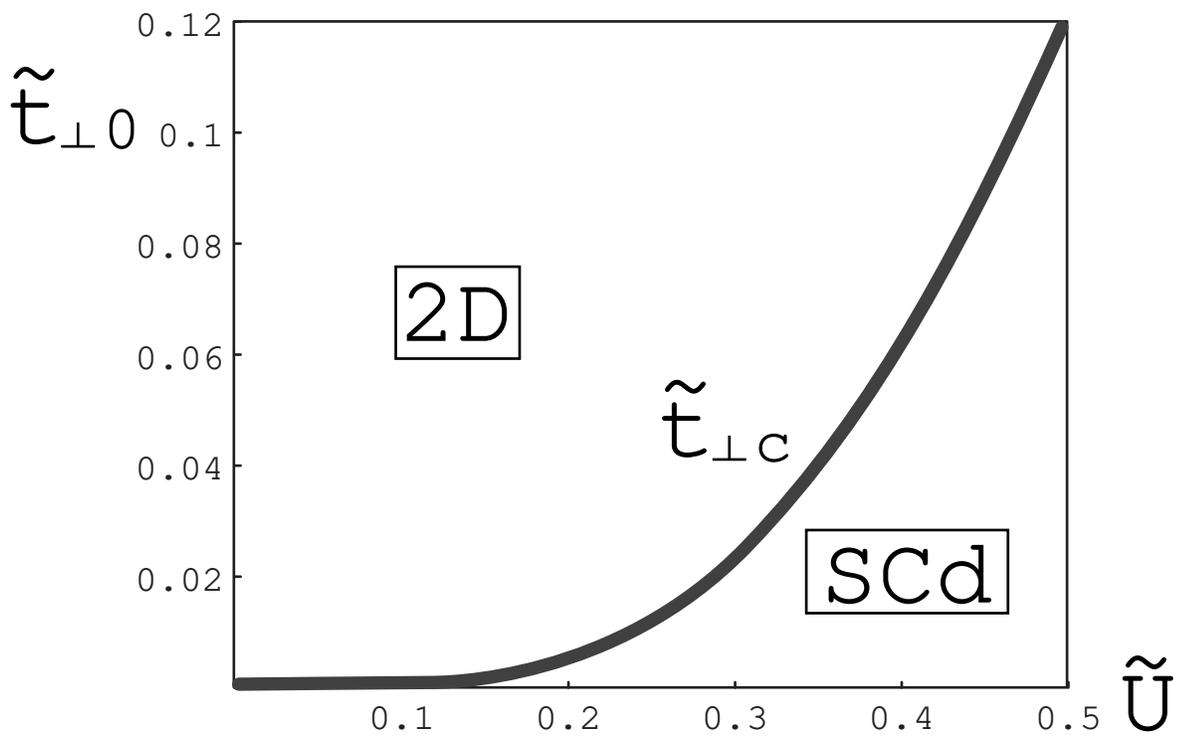
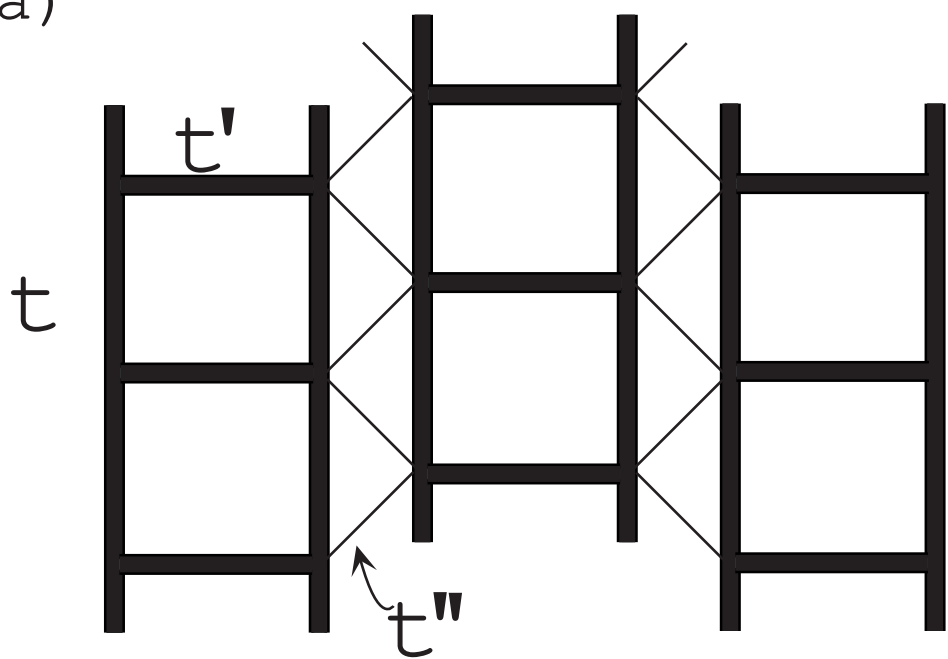
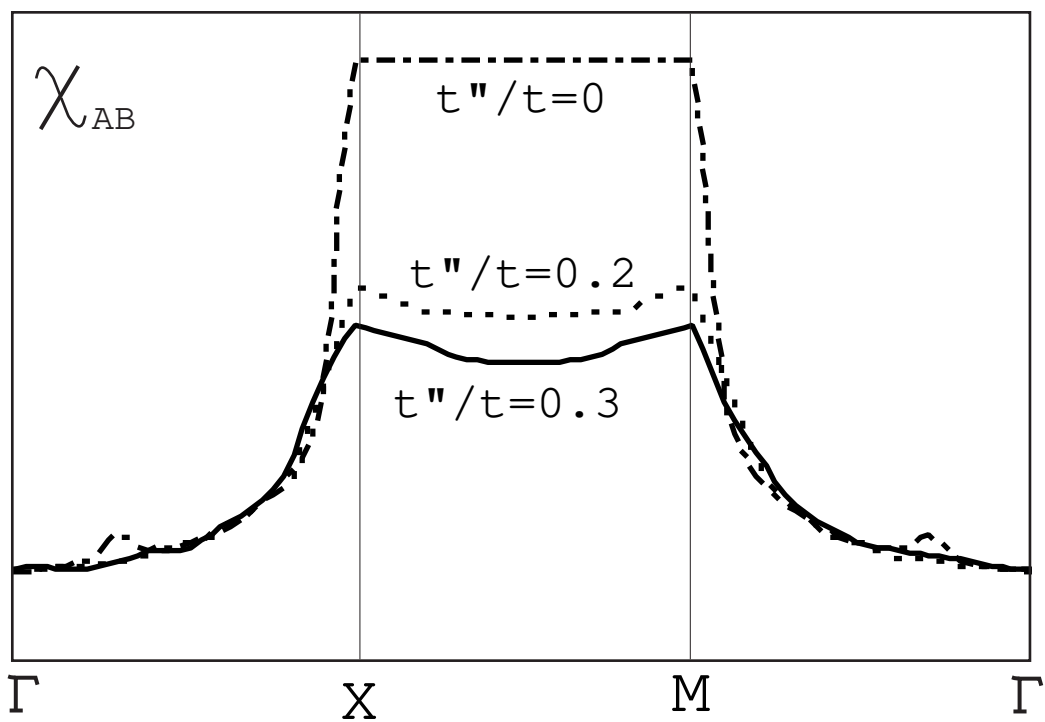
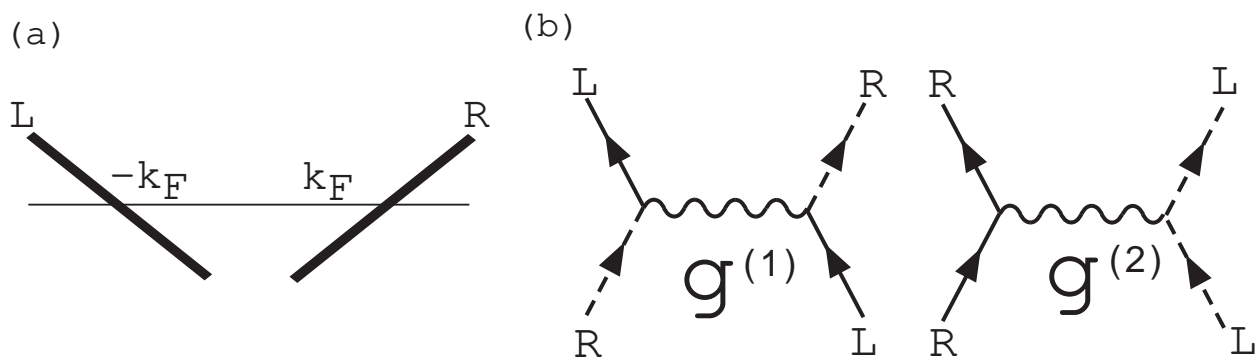
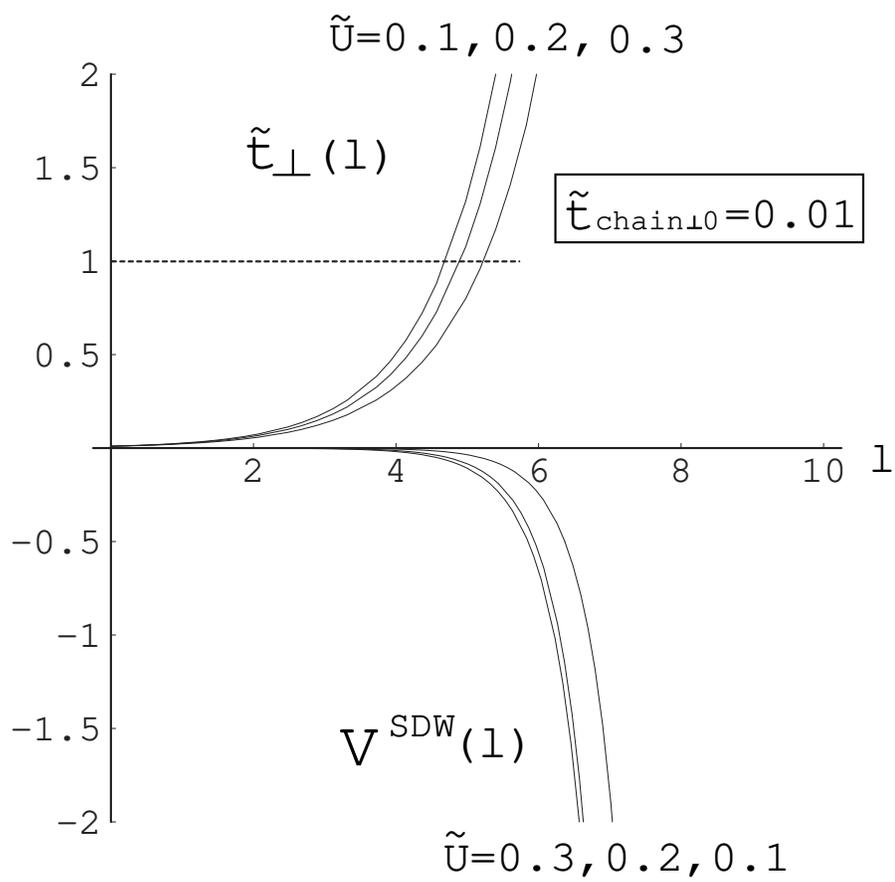
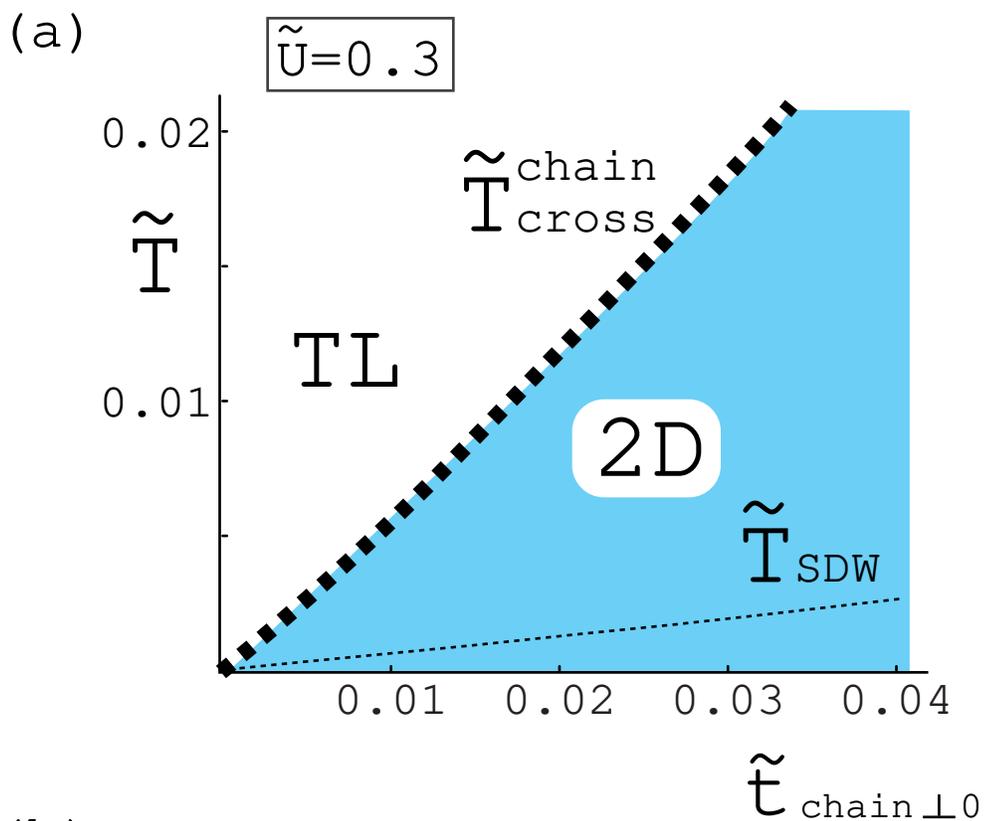
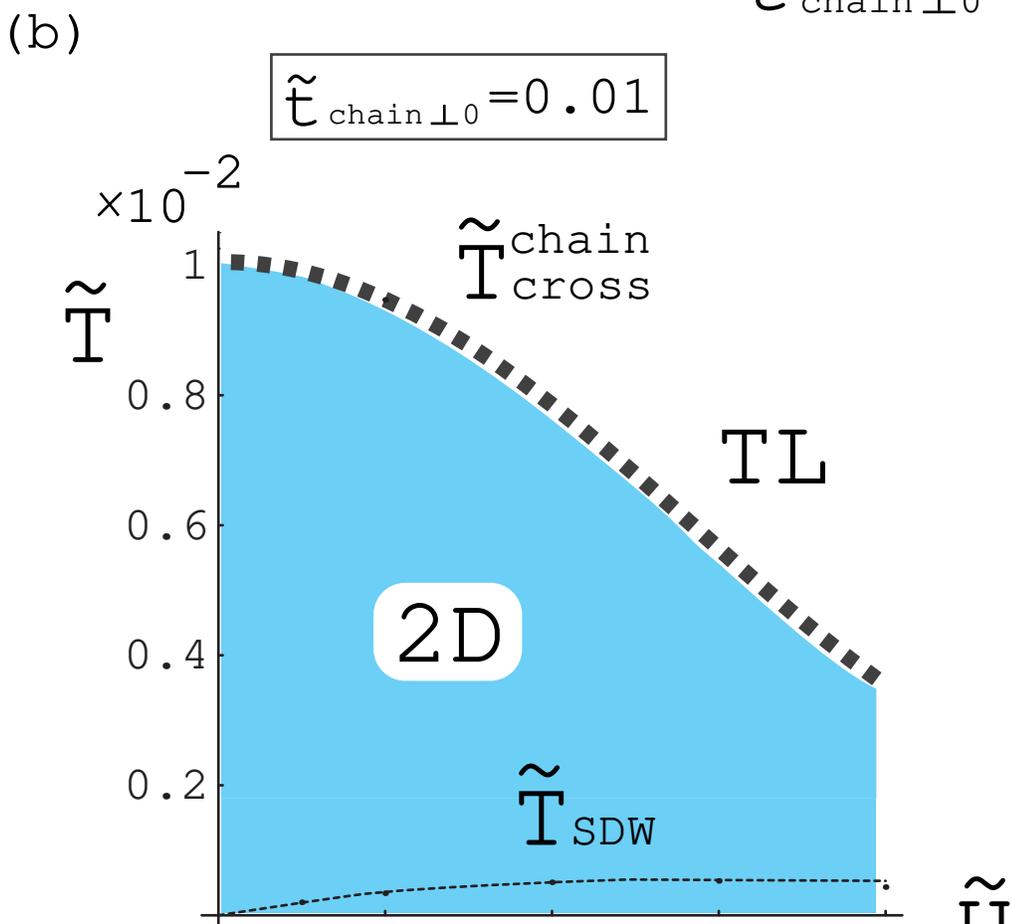
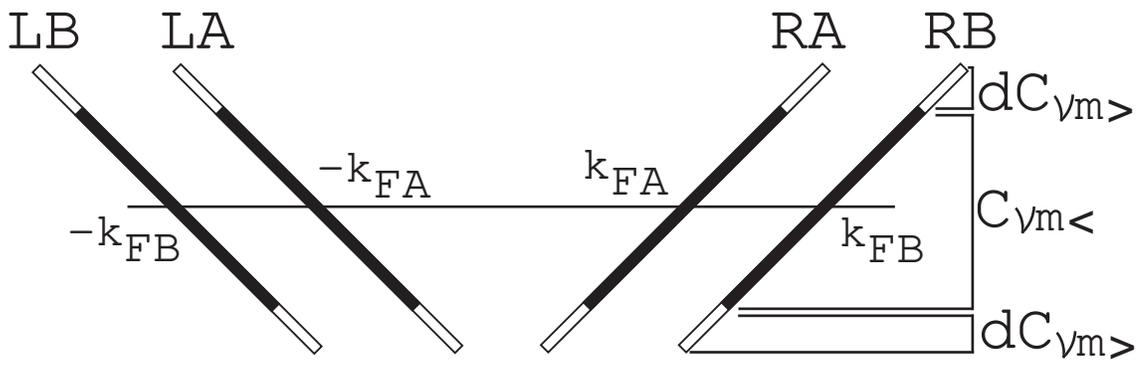
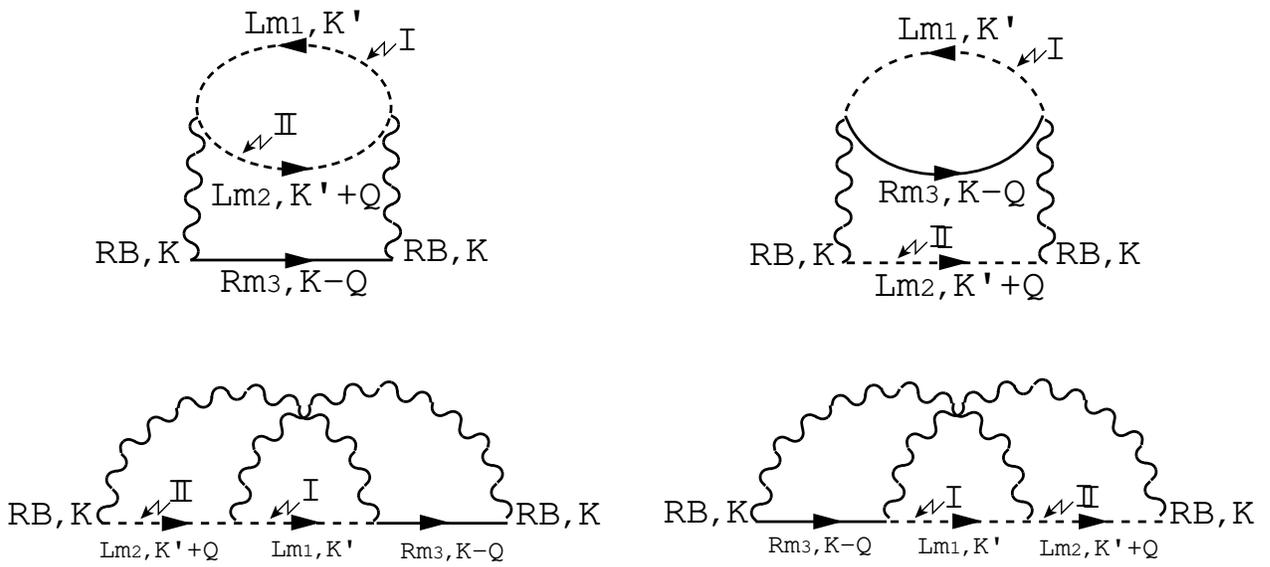
\end{document}